\documentclass[aps,prb,twocolumn,superscriptaddress, reprint]{revtex4-2}
\usepackage{blindtext}
\usepackage{appendix}
\usepackage{graphicx}
\usepackage{epsfig} 
\usepackage{caption}
\usepackage{soul}
\usepackage{subfloat}
\usepackage[justification=raggedright]{caption}
\usepackage[dvipsnames]{xcolor}
\usepackage[T1]{fontenc}
\usepackage{wrapfig}
\usepackage[english]{babel}
\usepackage{mathrsfs} 
\usepackage{amsmath}
\usepackage{amsfonts}
\usepackage{sidecap}
\usepackage{bbm}
\usepackage[multidot]{grffile}
\usepackage{subfig}
\usepackage{multirow}
\usepackage{url}
\usepackage[english]{babel}
\usepackage{siunitx}
\newcommand{\edits}[1]{\textcolor{black}{#1}}

\begin{document}

\title{Data-Driven Dynamical Mean-Field Theory: an error-correction approach to solve the quantum many-body problem using machine learning}

\author{Evan Sheridan$^*$}
\affiliation{Theory and Simulation of Condensed Matter, Department of Physics, King's College London, The Strand, London WC2R 2LS, UK.}
\author{Christopher Rhodes}
\affiliation{Theory and Simulation of Condensed Matter, Department of Physics, King's College London, The Strand, London WC2R 2LS, UK.}
\affiliation{AWE, Aldermaston, Reading, RG7 4PR, UK. }
\author{Francois Jamet}
\affiliation{National  Physical  Laboratory,  Teddington,  TW11  0LW, UK}
\author{Ivan Rungger}
\affiliation{National  Physical  Laboratory,  Teddington,  TW11  0LW, UK}
\author{Cedric Weber$^*$}
\affiliation{Theory and Simulation of Condensed Matter, Department of Physics, King's College London, The Strand, London WC2R 2LS, UK.}

\begin{abstract}

Machine learning opens new avenues for modelling correlated materials. Quantum embedding approaches, such as the dynamical mean-field theory (DMFT), provide corrections to first-principles calculations for strongly correlated materials, which are poorly described at lower levels of theory. Such embedding approaches are computationally demanding on classical computing architectures, and hence remain restricted to small systems, which limits the scope of applicability without exceptional computational resources. Here we outline a data-driven machine learning process for solving the Anderson Impurity Model (AIM) - the central  component of DMFT calculations. The key advance is the use of an ensemble error-correction approach to generate fast and accurate solutions of AIM. An example calculation of the Mott transition using DMFT in the single band Hubbard  model is given as an example of the technique, and is validated against the most accurate available method. This new approach is called data-driven dynamical mean-field theory (\textbf{d$^{3}$MFT}). 
\end{abstract}

\maketitle

\section*{Introduction} 

It is evident that many materials which are being used in emerging technologies, such as spintronics, quantum computing and high temperature superconductivity for example, are characterised by interacting  $d$ and $f$-band valence electrons situated at the boundary between being localised and delocalised \cite{Bhatti_MatTod_2017,Arute_Nature_2019,Zhou_NatRevPhys_2021}. Computational materials design is hampered by the complexity of these materials (their chemistry and structure) and their significant electron-electron interaction effects. This diminishes both calculation effort and understanding, and hampers technological development.   

In recent years significant progress in understanding the basic physics of strong electron correlation effects has been made by the dynamical mean-field theory (DMFT) method \cite{dmft_rev_mod}. Moreover, the harnessing of DMFT to mainstream materials modelling methods, such as density functional theory (DFT) to produce \edits{DFT}+DMFT theory, has generated material-specific capabilities which have begun to quantitatively address experimental contexts where correlation is a key determinant of material properties \cite{dftdmft}. 

Despite this, significant challenges remain when modelling real materials which feature many interacting electrons in multiple orbitals. At the heart of the LDA+DMFT method is the choice of quantum many-body solver for the Anderson Impurity Model (AIM). DMFT is based on a self-consistent mapping between local properties of a given material onto the AIM, a correlated impurity embedded in an infinite non-interacting bath. The AIM can be solved by high-level many-body methods such as quantum Monte Carlo \cite{dmft_rev_mod}, or its extension, the continuous-time Monte Carlo approach (CTQMC) \cite{A_Rubtsov_PRB_2005,P_Werner_PRL_2006}. The latter provides  an exact solution to the full rotationally-invariant AIM Hamiltonian within statistical error bars, but is limited to the imaginary time representation. The evaluation of real-frequency spectral quantities requires the use of ill-defined analytical continuation\cite{mem,Gunnarsson10}. Cluster extensions of DMFT which use multiple impurity sites suffer from the fermionic sign-problem when inter-orbital hybridizations are present. \edits{Many body perturbation theory, and for example its implementation via the iterative perturbation theory for DMFT, are a solid starting point for solving quantum many-body problems in the limit of small repulsion \cite{gabi_ipt}. They provide good agreement in this limit and can be extended to non-equilibrium and superconductivity.}

Other solvers are based on the numerical renormalization group (NRG) \cite{andrew_kondo_nrg}, which allow real axis calculations and access to Kondo physics, but remain challenging to extend for multi-orbital systems. Another method that provides solver solutions for real frequencies is the exact diagonalisation (ED) approach, where a finite size discretization of the AIM is used, through representation of the infinite bath in terms of a small number of effective \emph{bath-sites}. In typical implementations, the bath size ($N_b$) is restricted because of the exponential growth of the Hilbert space with the total number of sites $N_s$ (bath sites and impurity orbitals). Nonetheless,  Lanczos-based algorithms allows one to deal with large Hilbert spaces, where the discretization at low temperature \cite{lucaED,leibsch_ed_lanczos_dmft} is fine enough to compute observables accurately. Some success has been achieved by ED for multi-orbital systems with three or five orbitals  \cite{Capone02,ed_solver_liebsch_review,Ishida10}, but in general the limitation in the bath size limits the scope of applicability for realistically describing transition metal oxides, for example. To achieve of the largest number of bath sites, ED calculations have been extended to handle the single impurity embedding problem, allowing up to $\mathcal{O}(100)$ \cite{Arrigoni_2013,Lu_DMFT_2014,Ganahl_2015,Bauernfeind_2017} and $\mathcal{O}(300)$ \cite{Lu_2014} uncorrelated bath sites. Although the latter approaches generate approximations of the zero temperature Green’s function, the construction of systematically high energy excited states remains challenging. Clearly there is no single applicable AIM solution that is effective for all contexts and all parameters ranges, and  each method has its limitations. The stand-out technique is CTQMC, but it has poor low temperature scaling and is  computationally demanding - using it to address $f$-band materials, for example, demands exceptional computational facilities. This means high-throughput material design applications are beyond the reach of most researchers. 

The limitations of current solvers when addressing lanthanide and actinide materials, for example, has stimulated work on alternative solutions to the impurity problem. One conceptually different approach is an ensemble technique that draws upon the solutions of many approximate impurity solvers combined within a machine learning framework. Following the pioneering work of Arsenault \cite{Arsenault_PRB_2014_ES} and the work of Rigo and Mitchell \cite{mitchell_ai_1}, we use a neural network-based machine learning approach to DMFT and construct a training set of physically reasonable hybridization functions, and determine the impurity \edits{G}reen's functions corresponding to the training examples. The key feature of the quantum many-body solver presented here is that it uses a data-driven approach which focusses on training for the {\em prediction of errors} to calculate the Green's function rather than attempting to directly predict this quantity. The method we present here is \edits{dubbed as} "data-driven dynamical mean-field theory" - \textbf{d$^{3}$MFT}. To demonstrate the robustness of the \textbf{d$^{3}$MFT} method we give an example calculation of the Mott transition in a half-filled single band DMFT calculation for the Hubbard model on the Bethe lattice, and validate it against a highly accurate CTQMC solution.

The paper is organised as follows : In section I, we describe the machine learning approach for solving the Anderson Impurity Model.
In section II, we review the validation and benchmark of the \textbf{d$^{3}$MFT} approach. 
In section III, we apply the method to the Mott transition in the Hubbard model. 
Section IV provides our conclusions to this this work.
In Appendix A, we review the data augmentation strategy used for our 
machine learning based approach. 
In Appendix B, we describe the systematic study of the error scaling.
In Appendix C, we provide representative test samples from the quantum database.
In Appendix D, we discuss the training and validation loss of the neural network. 
In Appendix E, we discuss the inference of the trained neural network on representative candidate samples.

\section*{I. Method: Mapping the AIM to a Neural Network}

The AIM is a prototypical physical model that explains
quantum and thermal fluctuations in low-energy condensed phases of matter.
Anderson first used the model and applied it to the Kondo phenomena to explain
the increase of resistivity in metals at low temperature \cite{AIM_paper}. It
has also been heavily used in the field of heavy fermion systems
\cite{hewson_1993}. Decades after its initial formulation, it was given a new
application in the context of DMFT in an attempt
to provide a solution to the Hubbard model \cite{dmft_rev_mod}. It is often
the first port-of-call when going beyond the static mean field approach of DFT
to model strongly correlated electronic phenomena in real materials using LDA+DMFT methods
\cite{dftdmft}. Despite the resurgence of intense research efforts
focused on finding solutions to the AIM, they remain elusive. In particular,
analytical methods are few, and when successful are inconsistent
across the wide range of energy scales present in the model. Challenging parameter 
regimes are predominantly accessed by numerical methods
which follow either the Quantum Monte Carlo (QMC/CTQMC) or Exact Diagonalistion (ED)
methodologies. However, despite the conceptual and computational advances for
both of these techniques in the past decades there are still major limitations. 
These include - speed, accuracy at low temperatures,
number of correlated orbitals allowed, the fermionic sign problem - to name but
a few. As a result, there is considerable interest in developing
alternative impurity solvers that go beyond the current paradigm. Even so, a
solver that can be \emph{(i)} reliable across many parameter regimes and
\emph{(ii)} faster than ED or QMC would be a significant achievement, as it
would open the field of strongly correlated electron physics to high-throughput
techniques and \emph{ab-initio} molecular dynamics.

The transferability of machine learning and neural network ideas to 
quantum physics is shown in a collection of seminal works \cite{Parrinello, csanyi, kermode}, which
demonstrated that these methods can be used on \emph{ab-initio} derived 
databases to replace the computational bottleneck of DFT and Molecular Dynamics (MD).
In the last decade, a number of supervised and unsupervised methods
have been extended to many more problems in condensed matter physics, and in
particular to many-body systems. For example, the pioneering work of
Carrasquilla {\emph et al.} \cite{Carrasquilla2017} classified the paramagnetic to
ferromagnetic phase transition in the two dimensional Ising model using
data-driven methods alone, has instigated a number of supervised and unsupervised
methods that go beyond the classification of phases of matter such as encoding
the quantum many-body state with artificial neural networks
\cite{carleo,Carleo2018}. Moreover, the QMC method has benefited
in terms of speed \cite{qmcml1, qmcml2, qmcml3} and accuracy \cite{Broecker2017},
where the sign problem has been alleviated - using supervised learning
techniques. Another technique that has benefited from data-driven solutions is
the procedure of analytically continuing \cite{Arsenault_2017} a function of a
complex variable to the real axis. However, most relevant to the work
presented here paper is the application of machine learning methods to the AIM as in Ref. \cite{Arsenault_PRB_2014_ES}. Here it was shown that the Green's function of the AIM could
be learned using kernel ridge regression on a database of AIM solutions and their
physical parameters alone. Then in Ref. \cite{arsenault2015machine} the same
authors applied this method to solve the DMFT equations for the single band
Hubbard model which worked successfully in different limiting cases of the
localised and delocalised limits. More recently, neural networks combined with
different exact solvers have been used to generate larger databases of
solutions to the AIM \cite{sturm2020predicting, walker2020neural}.

In contrast to these developments, where the focus has been on building
models that relate the physical parameters of the AIM to its Green's function
solution, we propose a new framework where the model relates
{\em almost} exact solutions of the AIM with their {\em exact} solutions. 
To be specific, we use a neural network to learn the error between 
a family of solutions that are approximately correct, and what we
regard as exact solutions, \emph{i.e.} from Exact Diagonalisation. By using this approach we find that it retains the full many body features
present in the AIM across a wide range of model parameters. Another feature
of the work presented here is the automatic determination of a highly versatile
and symmetric basis choice for the impurity Green's functions that is used in
the machine learning procedure. We show that the computational resources needed
for a highly-accurate impurity solution with the machine learning-driven impurity solver are
much reduced, since all it requires for training data is a set of approximate 
solutions. Ultimately we incorporate the data-driven solvers into a DMFT calculation for the
Hubbard model for parameter regimes that span the Mott transition.

\subsection*{Outline of the machine learning protocol}
In this section we sketch the overall framework for creating a data-driven machine learning
solver for the AIM. To do this coherently we break the problem into three distinct units:

\emph{(i)} generating a database of solutions for the AIM \\
\emph{(ii)} training machine learning algorithms on the database to produce a
data-driven solver, and \\
\emph{(iii)} testing the efficacy of the data-driven solver. \\

An intermediate step (between step \emph{(i)} and step \emph{(ii)}) discusses some key 
data augmentations, parametrisations and transformations that 
enhance the performance of the neural network. 
This is presented in the Appendix A.

In what follows we address the single-band impurity problem, since there are reliable
approximate solutions to it and exact solutions can be
produced across a wide range of parameters with relative ease.

\subsection*{i) Database generation}

The construction of a high-quality database of training samples is of
paramount importance for any data-driven solution to a problem. Specifically,
there must be sufficient "good" samples that relate a collection of inputs to a
collection of outputs, such that after the training process a new set of input
instances will produce the most likely outputs, via the trained neural network. 

Before presenting the database construction details, we first need to view 
the AIM from a data-science perspective, rather than, a physics one. 
With this in mind, it is helpful to think of solving the AIM as a black box, 
and regard it only in terms of its inputs and outputs. In doing so, it 
will allow for an exact formulation its input samples $\mathbf{X}=\{ \mathbf{x_1}, \hdots, \mathbf{x_{N_S}}\}$ and output samples $\mathbf{Y}=\{ \mathbf{y_1}, \hdots, \mathbf{y_{N_S}}\}$, where $N_S$ is the number of database samples. 

The single-band AIM is completely
described by the following set of parameters $\{ U, W, \varepsilon, \beta, V\}$,
where $U$ is the Hubbard U parameter, $W$ is the half-bandwidth of the
bath-states, $\varepsilon$ is the impurity on-site energy, $\beta$ is the inverse
temperature and $V$ characterises the impurity-bath coupling. Furthermore, we
can break these contributions down for different sectors of the AIM Hilbert
Space, where $\{ U, \varepsilon \}$ represent the physics of the impurity while
$\{W, V\}$ represent the physics of the bath and $\beta$ is a property of the
entire system. In practice, these parameters are then passed to an impurity solver
and the Green's function of the impurity, in a specific basis
representation, is obtained as the solution of the AIM.

\begin{figure}[h!]
  \centering
  \includegraphics[width=\columnwidth]{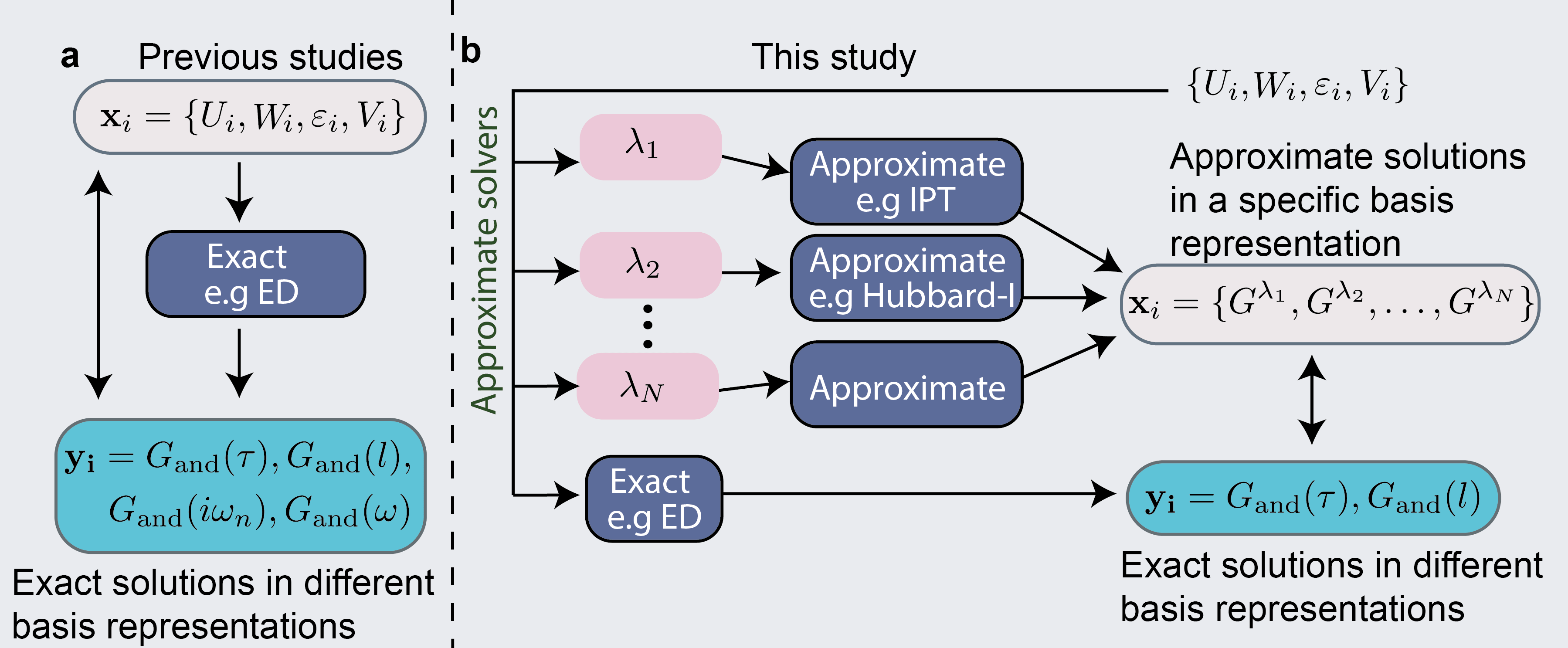}
  \caption[Machine learning the Anderson Impurity Model]{Depiction of two approaches for machine learning solutions of the Anderson Impurity Model. \textbf{a} The approach taken by \cite{Arsenault_PRB_2014_ES} which uses as input features $\mathbf{x}_{i}$ the physical parameters of the AIM $\{ U_i, W_i, \varepsilon_i, V_i\}$ \textbf{b} The approach taken her uses approximate solutions, expressed in different bases, $(\lambda_k)$, of the AIM as the feature vectors. } 
  \label{fig:AIM_approach}
\end{figure}

Indeed, in Ref.~\cite{Arsenault_PRB_2014_ES}, the authors used as their input features
$\mathbf{x}_{i} = \{ U_i, W_i, \varepsilon_i, V_i\}$ and $\mathbf{y}_i = 
\{G_{\textrm{and}}(\tau), G_{\textrm{and}}(l), G_{\textrm{and}}(i \omega_n), G_{\textrm{and}}(\omega) \}$ as the output vector, depending on the choice of basis. The basis choices above are the imaginary-time, Legendre, imaginary-frequency and real-frequency bases, respectively. In this case, each input vector has a
dimension of $4$ and each output vector has a dimension of $N_{b}$, where $N_b$ is the number of basis coefficients used to express the functional form of the solution. For example,
the number of imaginary-time points is usually $N_{\tau} > 200 $ to ensure
subsequent reliable interpolations, while it is known that the number of Legendre coefficients $N_{l} < 50 $ is more compact \cite{leg_poly_ref, Arsenault_PRB_2014_ES,qcqmc_ref}. In other words, this approach addresses the
question: given a set of AIM descriptors $\mathbf{x}_{i} = \{ U_i, W_i,
\varepsilon_i, V_i\}$ what is the functional form of its solution, which is basis dependent, \emph{i.e.}
$\mathbf{y}_i = 
\{G_{\textrm{and}}(\tau), G_{\textrm{and}}(l), G_{\textrm{and}}(i \omega_n), G_{\textrm{and}}(\omega) \}$ ? This procedure is illustrated in
Fig. \ref{fig:AIM_approach}a. In \cite{sturm2020predicting}, the
authors employ a similar approach for the input features, but attempt to learn
the spectral function $A(\omega)$ instead. In summary, while neglecting the
aspects of \emph{how} the learning is done and \emph{what} solvers are used to
construct the database of solutions, the approaches so far have tried to relate 
the physical parameters of the Anderson Impurity Model to its solutions, which are
defined in the imaginary-time, Legendre, imaginary- or real-frequency bases. 

The approach used here is different with regard to the input
vectors $\mathbf{x}_{i}$, where instead we use a combination of approximate
solutions to the AIM. We illustrate the idea
in Fig. \ref{fig:AIM_approach}b for the $i$'th AIM instance, which is still
characterised by the set $\{ U_i, W_i, \varepsilon_i, V_i\}$, but the feature
space increases by $N_{\lambda}N_{b}-4$, where $N_{\lambda}$ is the number of
approximate solvers used, where $\lambda$ refers to the different impurity
solver types, and $N_b$ is the number of basis coefficients used. For example, if we consider using two different approximate solvers $\{
\lambda_1, \lambda_2 \}$, then the input features for a model would be,
\begin{equation}
  \mathbf{x}_i = \{ G_{\text{and}}^{\lambda_1}(\tau), G_{\text{and}}^{\lambda_2}(\tau))  \text{ or }   \mathbf{x}_i = \{ G_{\text{and}}^{\lambda_1}(l), G_{\text{and}}^{\lambda_2}(l)) \},
  \label{eq:inputs_ml}
\end{equation}
in the imaginary-time or Legendre polynomial bases respectively, and the output vectors are given by,
\begin{equation}
  \mathbf{y}_i = G_{\text{and}}(\tau) \text{ or }  \mathbf{y}_i = G_{\text{and}}(l). 
  \label{eq:output_ml}
\end{equation}
In this case, the length of the feature vector is $2N_{b}$ and the length of
the output vector is $N_{b}$. Moreover, in lieu of trying to learn the
relationship between the input parameters $\{ U_i, W_i, \varepsilon_i, V_i\}$
and the exact solution of the AIM, we attempt to learn the {\em error} between a set
of approximate solutions of the AIM and its exact solution, generated using
Exact Diagonalisation. Equivalently, Quantum Monte Carlo can be used as the exact benchmark, but the computational resources required to generate exact solutions using this technique for the single-band Hubbard Model are generally more than those required by ED. We try to learn the error of a set of approximations, rather than using just one, noting that different 
approximate AIM solvers have their merits in different parts of the AIM parameter space. 

For instance, iterative perturbation theory is a well-known and successful diagrammatic method for solving quantum many-body problems in the weak-coupling limit, and can also capture some features of the strongly interacting limit \cite{gabi_ipt}. Similarily, the Hubbard-I approach is exact in the weakly hyrbisied atomic limit, but can qualitatively fail outside of this parameter regimes.
These approaches are quite often the first port of call in a many-body theorists arsenal when attacking a quantum many-body problem, due to their simplicity. While the weak-coupling expansion has its merits, for scenarios out at strong-coupling it can be qualitatively wrong. To address this, for example, the Non-Crossing Approximation (NCA) \cite{haule_nca} is the lowest order strong-coupling perturbative method that sums up all diagrams without crossing hybridisation lines, and like the previous approaches also only requires moderate computational resources. It is also possible to use basis-truncated approximate ED solutions that use less bath sites to represent the Weiss field, which thus provide a more consistent coverage of the parameter AIM space, albeit with larger errors when small numbers of bath fitting parameters are used. Thus, a natural extension for data driven methods lies with the combination of different approximations, which span a wider range of the AIM parameter space, rather than just using one quantum solver alone. 

Having established the form of the inputs and their associated outputs, we
now discuss the construction of the database. We first
deal with inverse temperature, $\beta$, and unlike the rest of the parameters, the Green's function
is explicitly dependent on it, so each database is constructed at a specific
$\beta$. The next parameters to decide are $\{ U, \varepsilon, D\}$, all of
which are randomly distributed between their extremal values. Therefore, for
each instance of an AIM, random samples of each are drawn from the uniform
distributions $U \in [U_{\text{min}},U_{\text{max}} ]$, $W \in
[W_{\text{min}},W_{\text{max}} ]$ and $\varepsilon \in
[\varepsilon_{\text{min}},\varepsilon_{\text{max}} ]$. 

Determining the hybridisation parameters is slightly more involved,
because of how it manifests itself in the impurity solver. The ED solver necessitates a discrete
representation of the bath parameters, while this is not true for other impurity
solvers. Therefore, for the overall versatility of the machine learning
protocol, we need to be able to deal with both discrete and continuous
representations of the bath. We first take on the continuous representation,
taking the Hilbert transform of specific form of the density of states, \emph{i.e.}
\begin{equation}
  G(z) = \int_{-\infty}^{\infty} \frac{A(\epsilon)}{z - \epsilon} d \epsilon.
  \label{eq:htrans}
\end{equation}
There are two choices in the implementation, the first is a semi-circular DOS given by,
\begin{equation}
  A(\epsilon) = \frac{2 \sqrt{-\epsilon^2}}{(\pi W)^2} \Theta(W - |\epsilon|), 
  \label{eq:semi_circ}
\end{equation}
and the second is a constant DOS, given by, 
\begin{equation}
  A(\epsilon) = \frac{\Theta(W^2 - \epsilon^2)}{2W}, 
  \label{eq:flat}
\end{equation}
where $\Theta$ is the Heaviside step-function. Using either truncates the limits
of integration in Eq. \ref{eq:htrans} from $-W$ to $+W$, hence $W$ is known as
the half-bandwidth. On the other hand, the discrete representation of the
hybridisation function is given by,
\begin{equation}
  \Delta(i \omega_n)  = \sum_{i=1}^{N} \frac{V_{i}^{2}}{i \omega_n - \epsilon_i}, 
  \label{eq:hyb_discrete}
\end{equation}
where $V_i$ and $\epsilon_i$ are the bath parameters, and
which become an additional parameter in the database construction. Specifically,
along with the number of samples in the database and the inverse temperature $\beta$, the number of bath
parameters determine the overall time it takes for the construction of the database.  After the number of bath sites is chosen, random samples of each are drawn from the uniform distributions $V \in
[V_{\text{min}}, V_{\text{max}} ]$ and $\epsilon \in [\epsilon_{\text{min}}, \epsilon_{\text{max}} ]$. 
To ensure that the discrete representation of the hybridisation function  retains consistent physical characteristic, its bath parameters are all scaled to the chosen values of $D$, such that both $\epsilon_i$ and $V_i$ are normalised by it and $\epsilon_i$ is centered on its weighted arithmetic mean with respect to $V_i^2$. Alternatively, it is possible to create discrete representations of the bath by treating $\epsilon_i$ and $V_i$ as fit parameters in Eq.~(\ref{eq:hyb_discrete}) to a continuous representation generated from the half-bandwidth $W$. 

\begin{figure}[h!]
  \centering
  \includegraphics[width=\columnwidth]{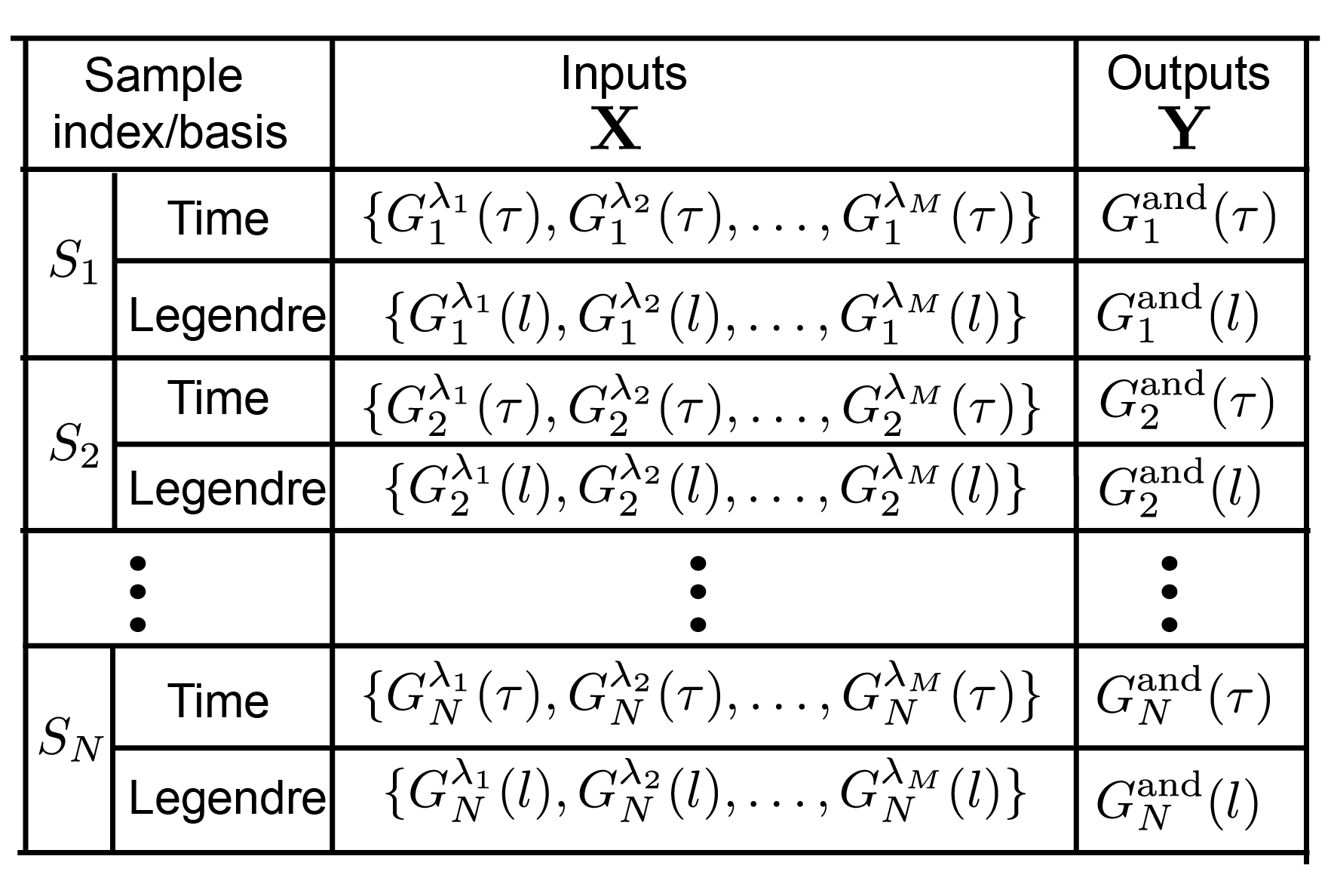}
  \caption[Schematic of the Anderson Impurity Model database layout]{ The database of single-band Anderson Impurity for M approximate ($\lambda_1, \lambda_2 , \hdots, \lambda_M$) solutions and exact solutions (from ED) for $S_N$ distinct $\{ U_i, W_i, \varepsilon_i, V_i\}$ parameter
    values, where $U$ is the correlation energy, $V$ is the hybridisation,
    $\varepsilon$ is the impurity filling and $W$ is the half-bandwidth in the bath.
    The database is categorised according to whether a solution is an input
    $\mathbf{X}$ or an output $\mathbf{Y}$ for a machine learning model. The
    Green's functions are shown to be expressed in the imaginary-time basis and Legendre basis,
    however they can also be expressed in different representations.}
  \label{fig:database_cartoon}
\end{figure}

The next step is to generate the database that will be used for the
training of the data-driven model. To do this, each instance of $\{ U_i, W_i,
\varepsilon_i, \Delta_i(i \omega_n)\}$ is passed to the set of approximate
solvers $\{\lambda_1, \hdots, \lambda_N \}$ as well as one exact solver. In this
case, the exact solution is obtained by the ED algorithm using a large number of
bath sites, generally between $4-6$ is enough to ensure a converged solution for the single-site AIM.
Fig. \ref{fig:database_cartoon} illustrates what a typical database looks like, where 
in this case there are $S_{N}$ samples that each use $M$ approximate solvers to 
generate the input data for the model. We also note that we have exclusively 
addressed the inputs and outputs of the database as being defined in the 
imaginary-time basis or Legendre polynomial basis, but this can be extended to any other basis 
representation (\emph{e.g} the real-frequency $\omega$ basis) with the appropriate transformation on each entry in the database. 

\subsection*{ii) Training a model solver}

\begin{figure}[h!]
  \centering
  \includegraphics[width=\columnwidth]{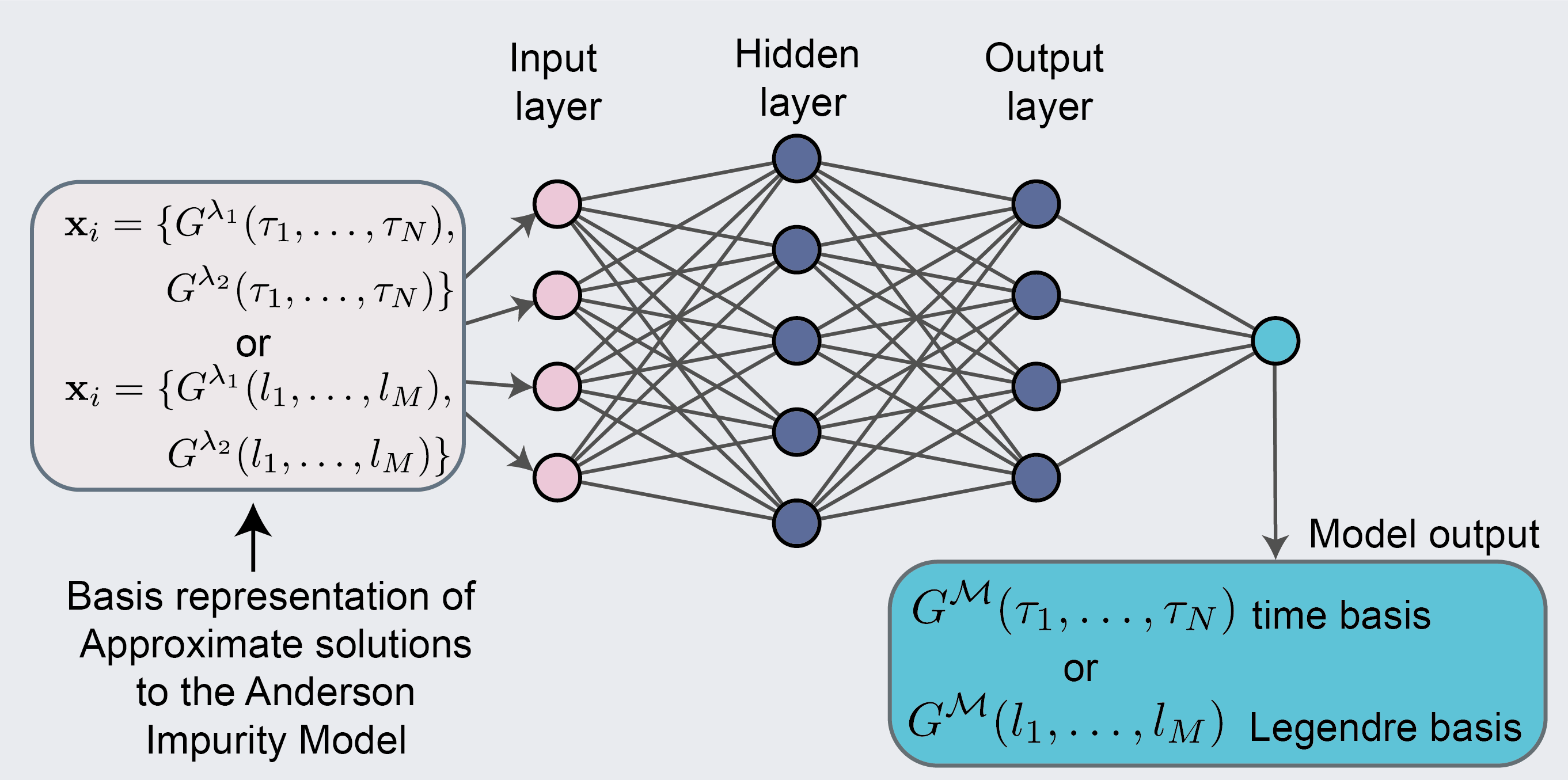}
  \caption{Depiction of the maximally connected feed-forward artificial neural network used to generate solutions of the Anderson Impurity Model to produce a model output $G^{\mathcal{M}}(\tau)$ or $G^{\mathcal{M}}(l)$. In the schematic, there is 1 hidden layer with 5 neurons and the output layer has 5 neurons.}
  \label{fig:NN}
\end{figure}

We now outline the details of the multivariate maximally connected neural network regression model that is used for
the training against the database we have just constructed. As established above, the set of
inputs for the model are $\mathbf{X} = \{\mathbf{x_{1}}, \hdots, \mathbf{x_{N_{s}}}\}$, where
$\mathbf{x_{i}}$ is a set of different approximate solutions of the AIM, while the
outputs are $\mathbf{Y} = \{\mathbf{y_{1}}, \hdots, \mathbf{y_{N_{s}}}\}$, where $\mathbf{y_{i}}$ is
an exact solution of the AIM given by ED. For all models trained in this section, they
proceed by minimising the cost function,
\begin{equation}
  C(\mathbf{X}, \mathbf{Y}, \boldsymbol{\alpha}) = \frac{1}{N_s}\sum_{j}^{N_s} [\mathbf{y}_{j}-g_{\boldsymbol{\alpha}}(\mathbf{x}_{j})]^{2},
  \label{eq:cost_function}
\end{equation}
with respect to the parameters $\alpha$ to produce a model
$g_{\boldsymbol{\alpha}}(x_{i}) := G^{\mathcal{M}}(x_{i})$, where
$G^{\mathcal{M}}(x_{i})$ is the model Green's function of the problem.
$G^{\mathcal{M}}(x_{i})$ is constructed such that the error between it and the
true solution $\mathbf{y}_i$ is minimised, and therefore
$G^{\mathcal{M}}(x_{i})$ corrects for the error between the approximate solution
$x_{i}$ and the exact one $y_{i}$, for all $N_s$ entries in the database. The neural 
network we use is shown as a schematic in Fig. \ref{fig:NN}. In the input layer, each neuron evaluates, 
\begin{equation}
  f(\mathbf{x.w+b}) = f \left( \sum_{i,j}w_{ij}G^{\lambda_j}(k_i) + b_1 \right)
  \label{eq:actv}
\end{equation}

with $f(\hdots)$ being the activation function of the input layer neurons
(coloured pink), index $i$ is associated to the feature (\emph{i.e} mesh point) and index $j$ indicates the approximate solver used, $w_{ij}$ are the set of neural weights, and $\mathbf{x}_i$ is in general of the format $\lambda_N$
entries, despite the depiction in Figure \ref{eq:actv} that suggests
that the number of approximate solvers is $2$. This procedure then repeats itself
as the values propagate forward through the network such that $f(\mathbf{x.w+b})$ of
each neuron are used as the inputs for the next layer in the
network, until eventually the output layer is reached. As the neural network is
being used to solve a regression problem, the output layer applies a linear
activation function to its neurons, which doesn't modify its input data. Therefore,
when the output layer is reached the cost function in
\ref{eq:cost_function} is evaluated for a "mini-batch" of samples, after
which the weights throughout the entire network are updated in accordance with
the backpropagation method. This procedure is then repeated
until $G^{\mathcal{M}}$ is found with weights $\alpha$ that minimise
$C(\mathbf{X}, \mathbf{Y}, \boldsymbol{\alpha})$.

It is important to keep in mind the number of parameters in the model
$g_{\alpha}(x_{i})$ so as to avoid potential over-fitting scenarios. For the neural network
presented in Fig. \ref{fig:NN} the total number of parameters $N_{\alpha}$ can be
determined by the following equation,
\begin{equation}
  N_{\alpha} = \sum_{l}(N_N^lN_N^{l-1}+1),
  \label{eq:nn_num_params}
\end{equation}
where $N_N^{l}$ are the number of neurons in layer $l$. For example, if there
are $20$ neurons in the input and hidden layer, $100$ neurons in the output layer and
$200$ is the length of the input vector, then the total number of parameters of
the network broken down per layer is given by:
\begin{equation}
  N_{\alpha} = \underbrace{20(200+1)}_\text{input layer} +   \underbrace{20(20+1)}_\text{hidden layer}  +  \underbrace{100(20+1)}_\text{output layer} = \underbrace{6540}_\text{total}.
  \label{eq:nn_num_params_example}
\end{equation}

The value of $N_{\alpha}$ is pertinent when considering sources of
data over-fitting, as it should not exceed to the total number of feature
observations in the database. In addition to what determines the number of
weights in the neural network, the the following series of adjustable
parameters have an effect on its performance - these are usually referred to as hyperparameters: 

\begin{itemize}
\item \textbf{Learning rate:}  step-size update for the weights of the network. 
\item \textbf{Mini-batch size:} the number of samples after which the neural network weights are updated. 
\item \textbf{Epochs:} the number of sweeps of the neural network. 
\item \textbf{Activation functions:} the family of non-linear neuron activation functions, including \emph{tanh}, elu and relu. 
\item \textbf{Cross-validation split:} the $\%$-split of the database between training/validation samples. 
\item \textbf{Basis functions:} the equidistant $\tau$-basis, adaptive $\tau$-basis or Legendre $G_l$-basis. 
\end{itemize}

Typically a hyperparameter grid search\cite{hp_search} is employed over
these parameters for a low value of the epoch number. In doing so, the model is
trained multiple times as it scans across different values of its parameters. Ultimately 
it will produce the final value of the cost function for both the
training and validation datasets, where the minimum will inform the final choice
of parameters to be used for a comprehensive training of the database.


\subsection*{iii) Physical applications}

\begin{figure}[h!]
  \centering
  \includegraphics[width=\columnwidth]{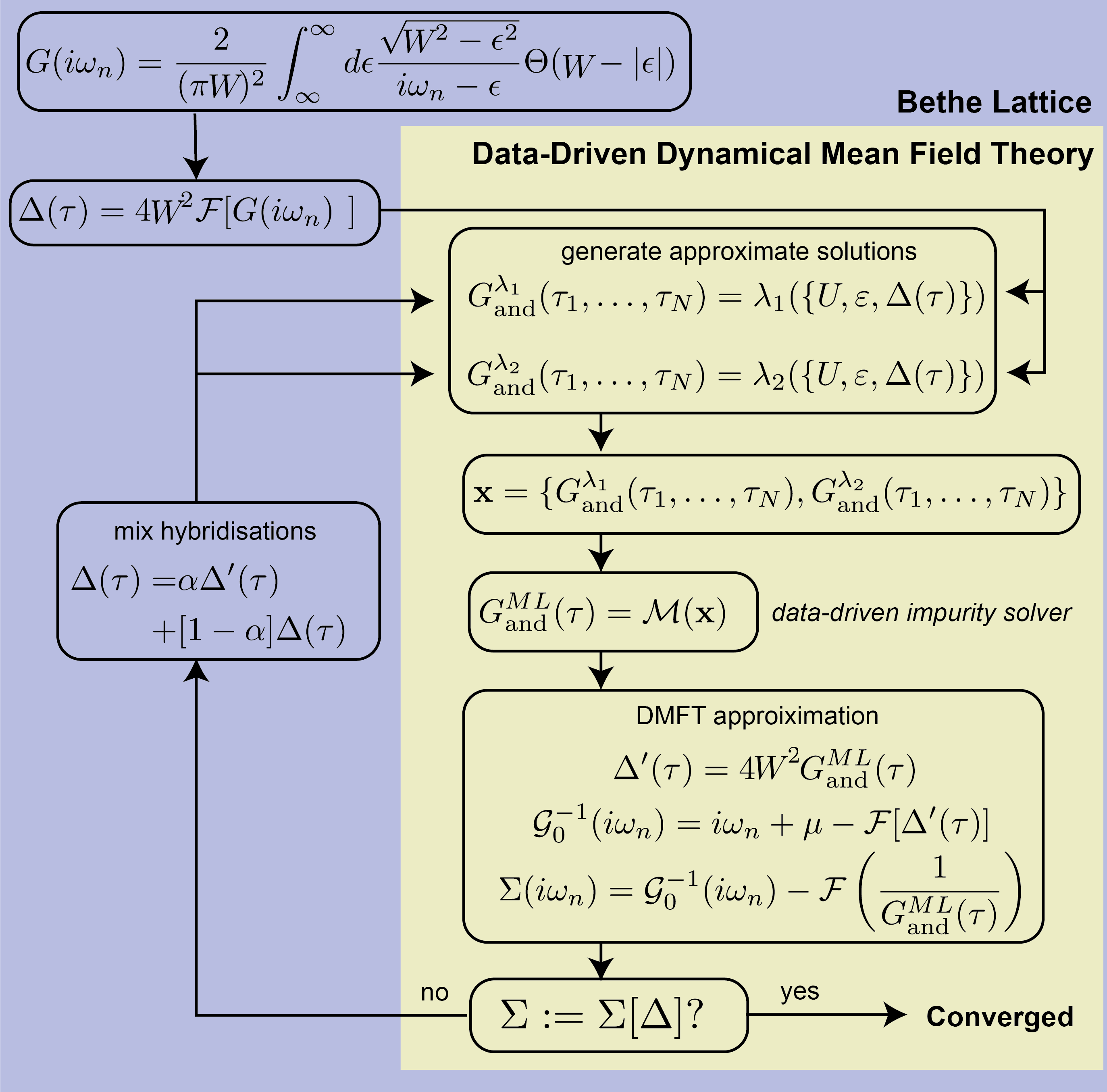}
  \caption[Data-driven Dynamical Mean Field Theory flowchart]{Flowchart of a data-driven DMFT calculation for a single band Hubbard model defined on the Bethe Lattice. The solver used is a data-driven impurity solver trained on approximate solutions of the Anderson Impurity Model. The colours separate the procedure by whether or not the calculation step falls inside the DMFT self-consistency cycle.}
  \label{fig:dmft_cartoon_LAIM}
\end{figure}

After training the neural network, we are left with a data-driven impurity solver for a
specific inverse temperature $\beta$. In Fig. \ref{fig:dmft_cartoon_LAIM} we
propose a workflow for solving the single-band Hubbard model on a Bethe lattice 
within the DMFT approximation using a data-driven model that has been trained (in this example) 
on two approximate solutions. To set up the problem, the local impurity
Green's function takes the form, 
\begin{equation}
  G(i \omega_n) = \frac{2}{(\pi W)^2}\int_{-\infty}^{\infty} d \epsilon \frac{\sqrt{W^2-\epsilon^2}}{i \omega_n - \epsilon} \Theta(W - |\epsilon|), 
  \label{eq:bethe_gf}
\end{equation}
where $W$ is the half-bandwidth of the system. After setting up the problem 
the only difference between this scheme and the
standard DMFT scheme \cite{dmft_rev_mod} is that instead of solving the AIM once
per DMFT iteration here it is solved twice with the two different approximate solvers ($\lambda_{1}$ and $\lambda_{2}$),
which are then used as inputs for the trained model $\mathcal{M}$. The
trained model then produces the impurity Green's function $G_{\text{and}}^{ML}(\tau)$
which is subsequently used to calculate the momentum independent self-energy $\Sigma(i
\omega_n)$. Additionally, a mixing parameter $\alpha$ is imposed when going from one iteration to the next,
\emph{i.e.} $\Delta(\tau) = \alpha \Delta'(\tau) + [1-\alpha] \Delta(\tau)$ to ensure that the 
self-consistency doesn't precipitately diverge from the previous iteration. By solving
this model, it is possible to analyse the Mott transition using the \textbf{d$^{3}$MFT} data-driven machine learning protocol. 

\section*{II. Benchmark and validation: A neural network for the AIM}

Here we present results that pertain to the \textbf{d$^{3}$MFT} framework outlined
above. We begin with a discussion  of the different aspects of the generated databases,
and this is followed by detail on the training of an artificial neural network with those
generated databases. We conclude by illustrating how the generated data-driven
solver is able to capture the Mott transition in the half-filled single-band Hubbard
model the using DMFT scheme presented above. 

\subsection*{Database of solutions for the Anderson Impurity Model}

Using the procedure outlined the previous section we generate a database of
size $N_{s}=10,000$ at inverse temperatures of $\beta=\{1, 2, 5, 10, 20, 50\}$ eV$^{-1}$ over the parameter
ranges indicated in Table \ref{tab:database_1} for discrete sets of bath
parameters. The range of temperatures chosen represent the high-temperature and intermediate temperature limits, whereby the features of the Green's function are smoother, and hence our choice of the range. Each database entry constitutes a random combination of all
parameters in Table \ref{tab:database_1}. The parameters chosen cover a range of
physical features, for example the Hubbard $U$ is uniformly randomly sampled
over the range $\{1, \hdots, 10\}$, in addition to $W \in \{1, \hdots, 10\}$ and
$\varepsilon \in \{-1, \hdots ,1\}$, then the various combinations of $U, W,
\varepsilon$ result in metals or insulators. Take for instance if $\{U, W,
\varepsilon\}=\{8,2,0\}$ then the result is insulating, and if $\{U, W,
\varepsilon\}=\{2,8,0\}$ the result is metallic. In Table  \ref{tab:database_1}
we clarify the notation for the Approximate solvers ED-[1,2,3]. ED-1 means solving the AIM with one bath site only, and hence results in a truncated approximation to the exact ED solution of the AIM (which in this case uses 4 bath sites). We justify the inclusion of these ED solvers since they are
exponentially faster than the true ED solution and are themselves approximate solutions, similar to IPT or Hubbard-I. 

\begin{table}[h!]
  \centering
  \begin{tabular}{| l | c | }
    \hline
    $U$ (eV) & $\{1, \hdots, 10\}$ \\ \hline
    $N_{\text{bath}}$, $\epsilon_i, V_i$& 4 \\ \hline
    $W$ (eV) &  $\{1, \hdots, 10\}$ \\ \hline
    $\varepsilon$ & $\{-1, \hdots ,1\}$ \\ \hline
    $\beta$ (eV$^{-1}$) & $\{1, 2, 5, 10, 20, 50 \}$ \\ \hline
    $N_{\text{samples}}$ & 10,000 \\ \hline
    $\mathcal{S}$ & Hubbard-I, IPT, NCA, ED-[1,2,3] \\
    \hline
    \end{tabular}
    \caption[Parameter selection for the database of Anderson Impurity Model solutions]{Parameter selection for the database of AIM solutions shown in Fig. \ref{fig:database_generation} where $\{p_i,\hdots,p_f\}$ denotes that a parameter is randomly selected from this interval $[p_{i}, p_{f}]$. $U$ is the Hubbard interaction,
    $N_{\text{bath}}$ stands for  the number of bath sites, $W$ is the Half-bandwidth, $\varepsilon$ is the electron on-site energy, $\beta$ is 
    the inverse temperature, $N_{\text{samples}}$ is the number of database entries, and $\mathcal{S}$ denotes the total ensemble of approximate quantum solvers used in the ML approach. ED-[1,2,3] denotes the exact diagonalisation solver with respectively $1,2,3$ bath sites.
    }
    \label{tab:database_1}
\end{table}

\begin{figure}[h!]
  \centering
  \includegraphics[width=\columnwidth]{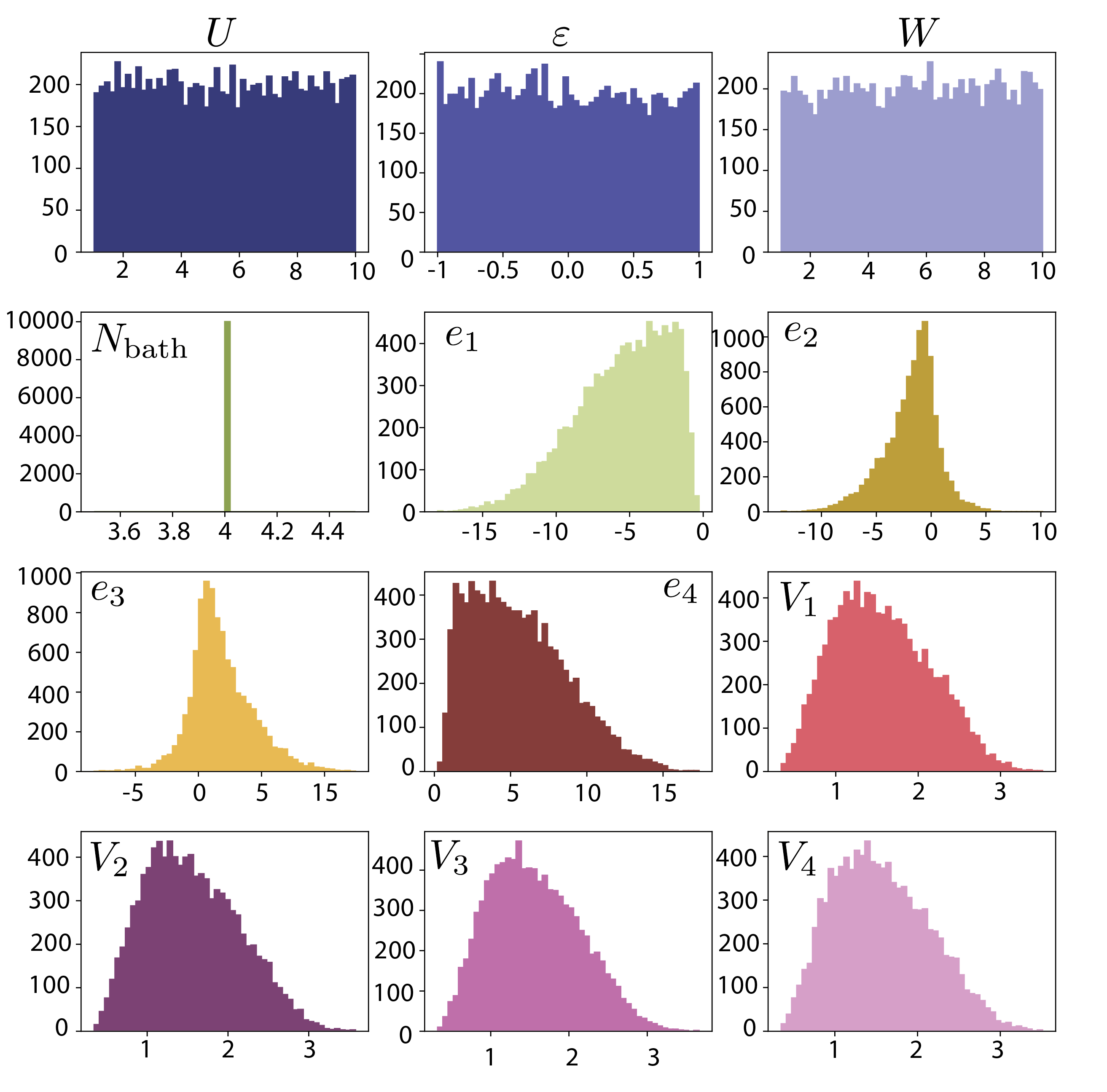}
  \caption[Distribution of Anderson Impurity Model parameters]{Typical statistical distribution of the Anderson Impurity Model parameter space for a database for $\beta=20$ eV$^{-1}$, $N_{bath}=4$ with 10,000 entries.}
  \label{fig:database_generation}
\end{figure}

Furthermore, in Fig. \ref{fig:database_generation} we show the distribution
of all chosen parameters for the $10,000$ samples in the database corresponding to
$\beta=20$ eV$^{-1}$. As expected, $\{U, W, \varepsilon\}$ are distributed
evenly, $N_b=4$ is constant as the number of bath-sites is not changed, and $\{
\epsilon_i, V_i \}$ are chosen by normalising to $W$. While the illustrated database is not
the only one that could be considered, it is not a special choice.  For all other databases we analysed the distribution of parameters behaves similarly to the $\beta = 20$ eV$^{-1}$ case presented.  

In Appendix C, we review the strength and weaknesses of the Hubbard-I, IPT and NCA solvers for representative samples of the database against the corresponding exact diagonalisation results using the solver of Ref. \cite{cedric_solver_ES} at tempartures $\beta=\{1,20,50\}$ eV$^{-1}$ shown in Figure~(\ref{fig:database_gfs}). The latter provide valid approximations of the AIM in different limits (Hubbard-I and IPT are good in the weakly hybridised limit, NCA is a good  approximation for stronger interactions). In general, the Hubbard-I, IPT and NCA solvers are however in quantitative and qualitative disagreement. For example, the Hubbard-I solver indicates for highly hybridised AIM an insulating solution when in reality the system is metallic. Nevertheless, we emphasise that this behaviour is expected and welcomed, since the end-goal is to systematically correct for the error between the approximate and exact solutions. A systematic analysis of the RMSD error for each of these solvers with respect to the true solution is also reported in Appendix B. This analysis reveals some of the in depth features of the approximate solutions, and highlights potential parameter choices that the network might struggle with.

\subsection*{A Neural Network Impurity Solver}

The very first step to training a machine learning model is the hyperparameter
grid-search over its independent parameters using tensorflow \cite{tensorflow2015-whitepaper}. Specifically, for our neural network we coarse-grain the number of epochs to $20$, set the inverse temperature
to $\beta = 1$ eV$^{-1}$ and scan across all combinations of parameters in Table 
\ref{tab:hyperparameter_search}. Ultimately, $401$ different neural
networks are trained and the combination of parameters with the minimum cost
function $\sim10^{-6}$ is given by that combination of parameters shown in the
"Optima" column of Table \ref{tab:hyperparameter_search}. Additional
fixed parameters in the grid search are: evenly spaced time grid, Hubbard-I, IPT and NCA solvers
as inputs $\mathbf{x}_i$ to the neural network as they all require minimal computational resources in comparison to the ED methods, no data augmentation, and the
$\mathcal{T}_4$ imaginary-time transformation from
Fig. \ref{fig:trans_table}. It is noteworthy that either increasing the complexity
of the network, \emph{i.e.} increasing its depth beyond $1$, or increasing the
learning rate to an order beyond $10^{-2}$ has detrimental effects on the
minimisation of the cost function. Practically, it would be computationally
prohibitive to perform this grid search for every $\beta$ and their additional
free-parameters. In what follows, all networks use the optimal values as
specified in Table \ref{tab:hyperparameter_search} and use an $80/20$
cross-validation split, \emph{i.e.} $80\%$ training data and $20\%$ validation
data.

\begin{table}[h!]
  \centering
  \begin{tabular}{| c | c | c |}
    \hline
    Hyperparameter & Range & Optima  \\ \hline 
    $\eta$ & $[ 0.0001, 0.0002, 0.001, 0.01 ]$ &  0.0002 \\ \hline
    Mini-batch size & $\{4, 8, 16, 32, 64\}$ & 8 \\ \hline
    Optimiser &$\{$Adamax, Nadam$\}$ & Nadam \\ \hline
    Activation functions &$\{$elu, relu, \emph{tanh} $\}$ & \emph{tanh} \\ \hline
    Hidden Layers & $\{1, 2, 3, 4\}$ &  1   \\ \hline    
    \end{tabular}
    \caption[Hyperparameter grid-search specifications]{Hyperparameter grid-search over the neural network parameters with a fixed number epochs = 20 and $\beta=1$ eV$^{-1}$}
    \label{tab:hyperparameter_search}
\end{table}

With the networks' hyperparameters chosen, we now outline how the various models
$\mathcal{M}_i$ are constructed and are used to test the best set
of approximate $\mathbf{x}_{i}$ inputs to produce the minimal cost function. These
models are differentiated by their input parameters as tabulated in Table 
\ref{tab:nn_train}. For example, $\mathcal{M}_1$ consists of the IPT, NCA and Hubbard-I solvers as inputs, while $\mathcal{M}_3$ combines many
more. If each approximate input has $N_{b}$ mesh points then the feature
space for $\mathcal{M}_1$ is $3N_{b}$, while for $\mathcal{M}_3$ it is
$6N_{b}$. As a result, the number of weights in the network for
$\mathcal{M}_3$ will be greater than for $\mathcal{M}_1$.

\begin{table}[h!]
  \centering
  \begin{tabular}{| c | l | }
    \hline
    Model & $\mathbf{x}_i$  \\ \hline 
    $\mathcal{M}_1$ & [Hubbard-I, IPT, NCA] \\ \hline
    $\mathcal{M}_2$ & [IPT, NCA, ED-2, ED-3] \\ \hline
    $\mathcal{M}_3$ & [IPT, NCA, ED-1, ED-2, ED-3, Hubbard-I] \\ \hline
    $\mathcal{M}_4$ & [ED-2, ED-3] \\ \hline
    $\mathcal{M}_5$ & [IPT, ED-3] \\ \hline
    $\mathcal{M}_6$ & [NCA, ED-3] \\ 
    \hline
    \end{tabular}
    \caption[Different types of neural network models considered for training]{The 6 different $\mathcal{M}_i$ input data models considered for training. Each model differs only in its combination of solver ensembles.}
    \label{tab:nn_train}
\end{table}

\begin{figure}[h!]
  \centering
    \includegraphics[width=\columnwidth]{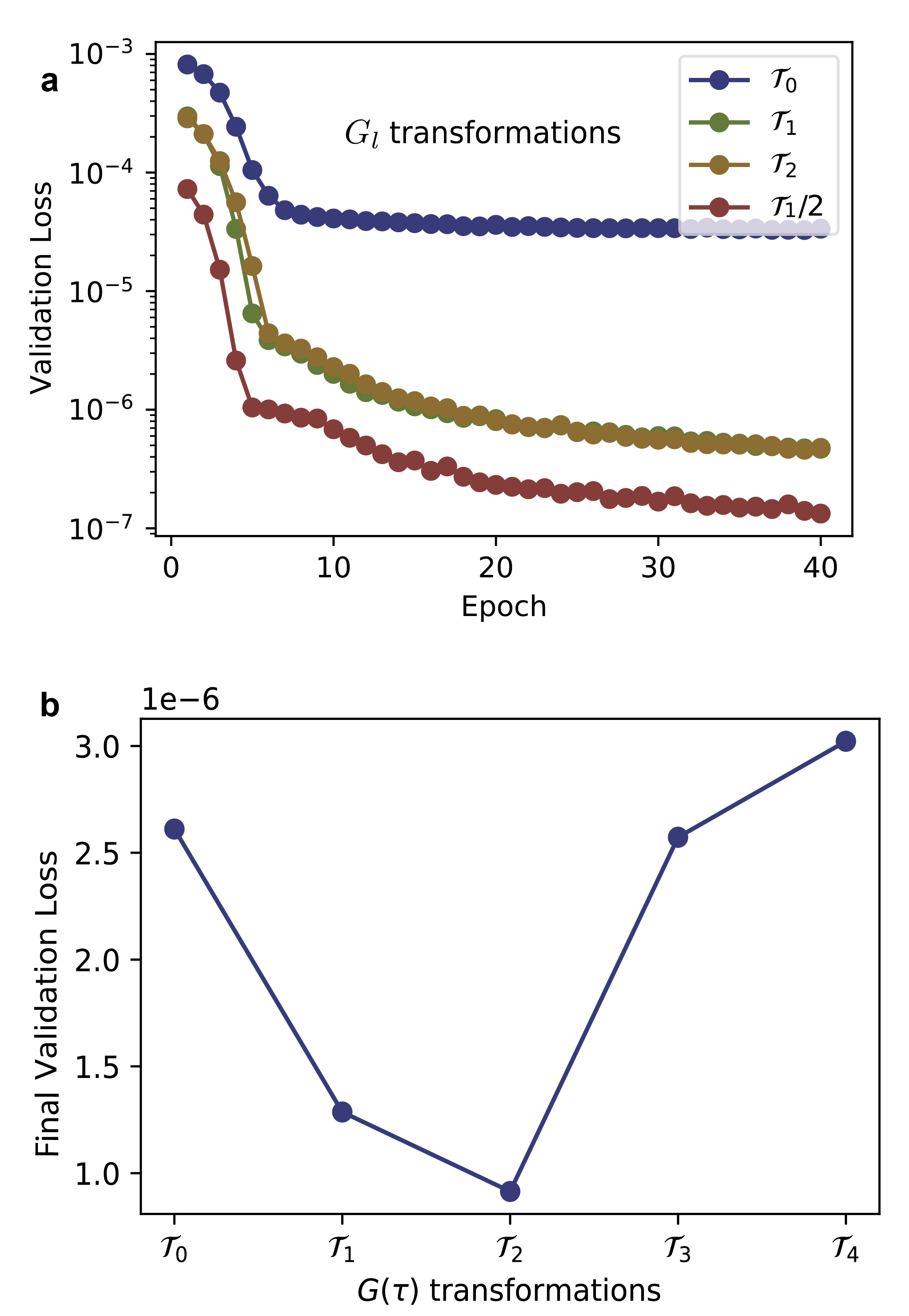}
  \caption[Validation loss for different $G(\tau)$ transformations]{\textbf{a} The $\beta=1$ eV$^{-1}$ validation loss for the Legendre basis transformations in Fig. \ref{fig:trans_table}b \textbf{b} The final value of the validation loss after 40 epochs at $\beta=1$ eV$^{-1}$ for the imaginary-time basis transformations shown in in Fig. \ref{fig:trans_table}a}
  \label{fig:res_transforms}
\end{figure}

In the Appendix A we propose a collection of data scaling
transformations of the input and output data which improve the training of the neural network in the
imaginary-time and Legendre bases. Fig. \ref{fig:res_transforms} presents
the validation loss for these scenarios, for $\beta = 1$ eV$^{-1}$ and model type
$\mathcal{M}_1$. For the Legendre basis the effect of data
transformations is quite siginificant, as shown in Fig. \ref{fig:res_transforms}a.
Here we see that by applying a \emph{tanh}-type Legendre transformations that the final 
value of the loss can be improved by at least 2 orders of magnitude, reduced from $10^{-4}$ to 
$10^{-6}$. We expect the effect of this transformation to by enhanced for larger values of $\beta$ (lower temperatures), where the range of $G_l$ can greatly exceed the value of unity, and therefore necessitates the application of an appropriate data transformation. At higher temperatures (\emph{i.e} lower $\beta$), the Legendre coefficients are often bounded close to unity, and so the neural network is less sensitive to the untransformed input as compared to lower temperatures. 

\begin{figure}[h!]
  \centering
  \includegraphics[width=\columnwidth]{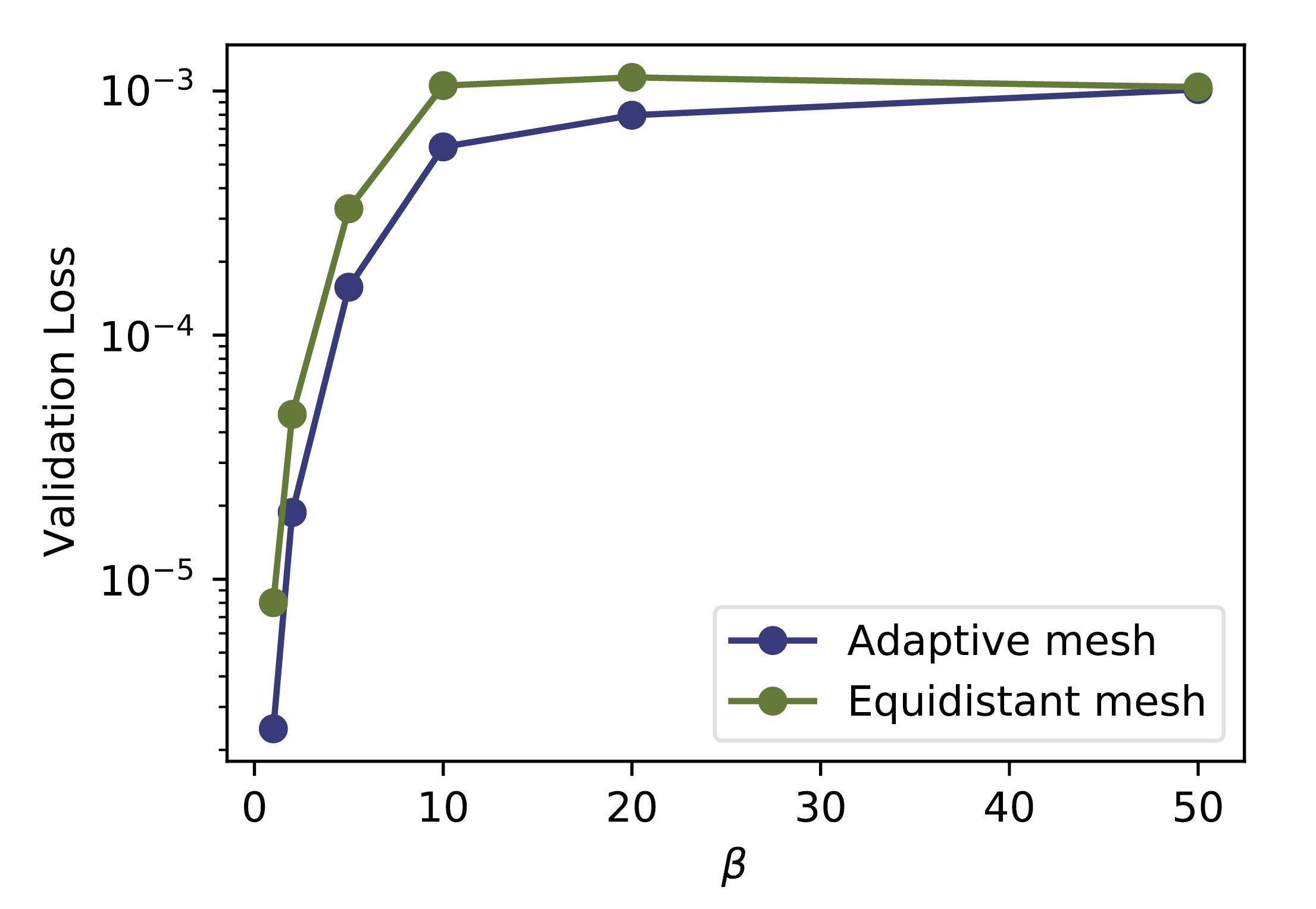}
  \caption[Validation loss for different $\tau$ meshes]{The value of the validation loss after 30 epochs for different values of inverse temperature $\beta$ in the imaginary-time basis with equidistant and adaptive mesh spacings.}
  \label{fig:ad_vs_eq}
\end{figure}

We show the final value of the validation loss after $30$ epochs for the
equidistant imaginary-time basis in Fig. \ref{fig:res_transforms}b. For $\beta=1$ eV$^{-1}$ the converged values are all within $10^{-6}$ of each other,
indicating a minimal impact of this transformation scheme at low
values of $\beta$. Hereafter, we use the $\mathcal{T}_2$ transformation in
imaginary-time and $\mathcal{T}_1/2$ for the Legendre basis. The hypothesis that
as $\beta$ increases, \emph{i.e.} temperature falls, the network becomes 
harder to train is affirmed, indicating that a deterioration of the 
error landscape shown in Fig. \ref{fig:error_impurity_solver}d and Fig. \ref{fig:error_impurity_solver_NCA}d
is manifested in the predictive qualities of the network. Notably, using an
adaptive $\tau$-mesh over an equidistant one is more effective in improving
the quality of the network at high temperatures, as Fig. \ref{fig:ad_vs_eq} illustrated for
different values of $\beta$ at the converged validation loss value. However, for
$\beta = 50$ eV$^{-1}$ there is a degeneracy between the equidistant and adaptive
meshes. We attribute this to the fact that increasing values of $\beta$
necessitates an increase in the number $N_{\tau}$ points in the $[0,\beta]$
interval to capture the emergence of more features near the boundaries. Since the total number of imaginary-time points, 
$N_{\tau}$, is fixed to $153$, it is not alarming to observe such a result.
Production quality data on larger values of $\beta$ therefore requires more
$N_{\tau}$ points, which is similar to the requirement of increasing the
number of $N_{l}$ Legendre coefficients needed at higher temperatures. Unless otherwise stated, all subsequent imaginary time based neural networks are trained on adaptive meshes to circumvent the issues of equidistant sampling in the stuided temperature regimes. 

\begin{figure}[h!]
  \centering
\includegraphics[width=\columnwidth]{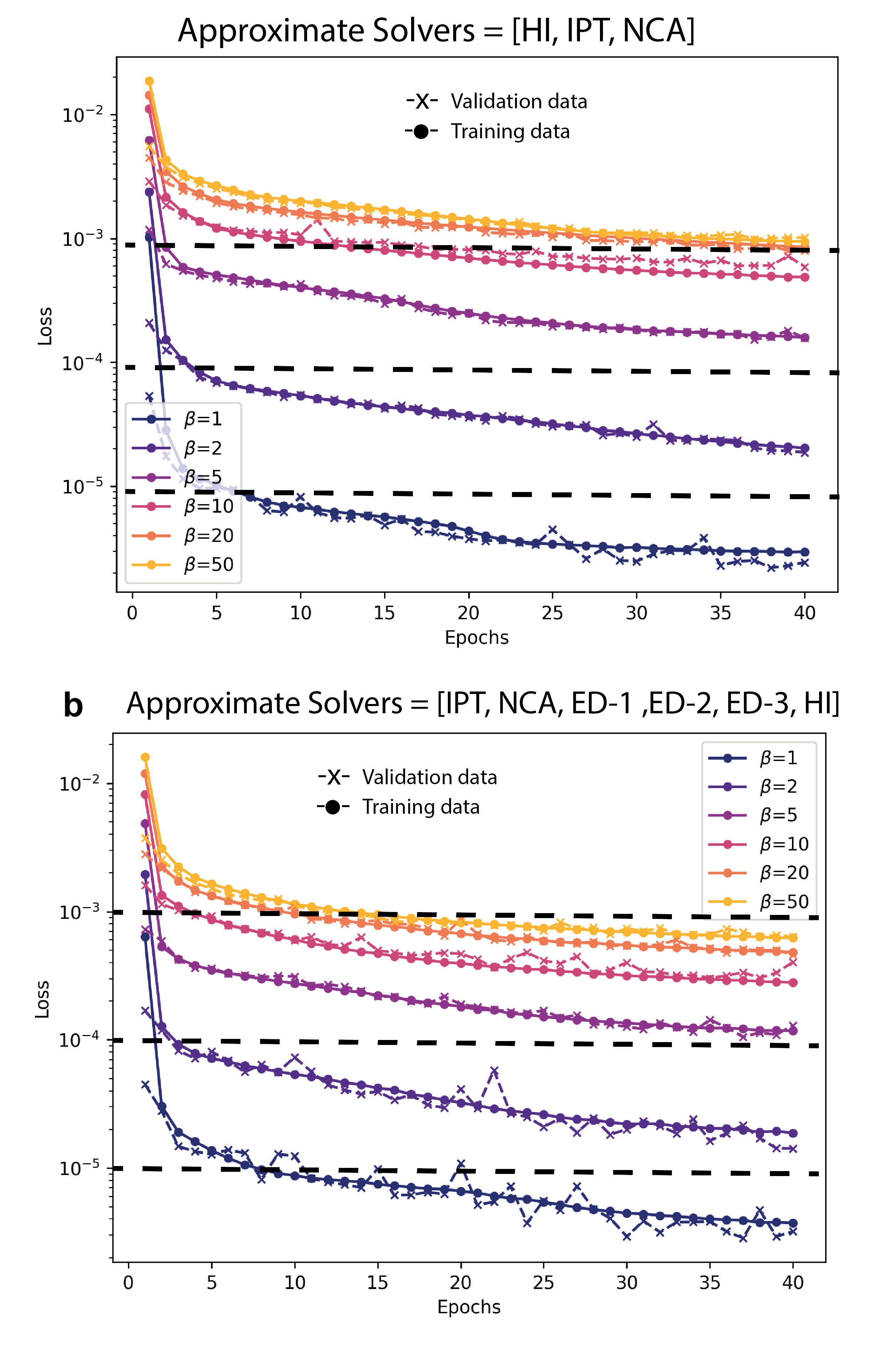}
  \caption[Cost function of the Neural Network for different approximate solvers]{Cost function for the adapative $\tau$ mesh for different inverse temperatures $\beta$ using as input to the neural network the solvers \textbf{a} [Hubbard-I, IPT, NCA] \textbf{b} [IPT, NCA, ED-1, ED-2, ED-3, Hubbard-I]. \edits{IPT is the Iterated Perturbation Theory Solver, NCA is the Non Crossing Approximation solver and ED-[1,2,3] are the truncated ED solvers.}}
  \label{fig:loss_adaptive}
\end{figure}

To prevent the scenario of over-fitting, we next analyse the minimisation of the
loss and account for the behaviour of both the training and validation sets. In
Fig. \ref{fig:loss_adaptive}a-b we show exactly this, across a number of
inverse temperatures, for models $\mathcal{M}_1$ and $\mathcal{M}_3$ in the imaginary-time basis. As the
validation loss is in lockstep with the training loss we conclude that neither
models are over-fitted. In addition, it is also very clear from these plots that
changing the form of the input does not change the fact that as $\beta$
increases that the network becomes harder to train. While it can be seen that
$\mathcal{M}_3$ outperforms $\mathcal{M}_1$ for larger values of $\beta$, this
improvement is negligible in practice. Furthermore, the variation in the
training data actually increases as more approximate solvers are added, as can
be seen for $\beta = 50$ eV$^{-1}$ in Fig. \ref{fig:loss_adaptive}. Therefore, the
training process is more sensitively dependent on the quality of the approximate
solvers, rather than their quantity.

\begin{figure}[h!]
  \centering
  \includegraphics[width=\columnwidth]{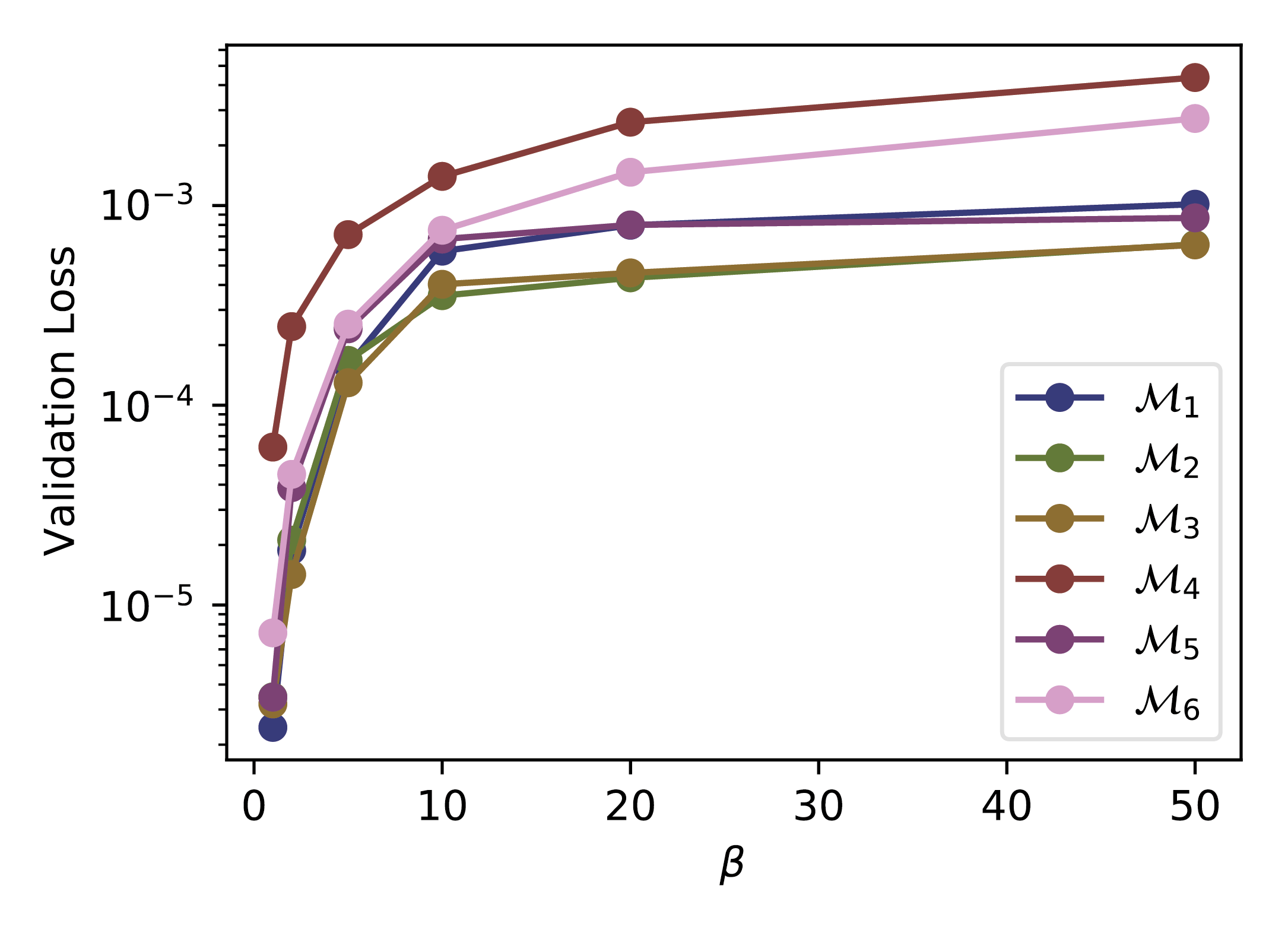}
  \caption[Final value of the loss function for different inverse temperatures $\beta$ and $G(\tau)$ transformations]{Final value of the loss function for different inverse temperatures $\beta$ in the adapative imaginary-time basis after 40 epochs for different variations of the model $\mathcal{M}$ input parameters, specified in Table \ref{tab:nn_train}. }
    \label{fig:final_valid_loss}
\end{figure}

To assess this further, in Fig. \ref{fig:final_valid_loss} we show the
validation loss function after $40$ epochs at different values of $\beta$ for all
models $\mathcal{M}_i$ in Table \ref{tab:nn_train}. The efficacy of the IPT solver is shown here, as all models that do not include it
under-perform relative to it. In particular $\mathcal{M}_4$ and $\mathcal{M}_6$,
which both include the NCA solver (however IPT is absent), are
consistently the worst performers. At small values of $\beta$ it is possible to
use any of the models and still achieve an accuracy of up to $10^{-5}$, however
as $\beta$ increases, the inclusion of the IPT accounts in some
cases for an order of magnitude improvement in the loss. The full training and validation loss functions for the additional models studied in Table \ref{tab:nn_train} are in Appendix D. 

\begin{figure}
\includegraphics[width=\columnwidth]{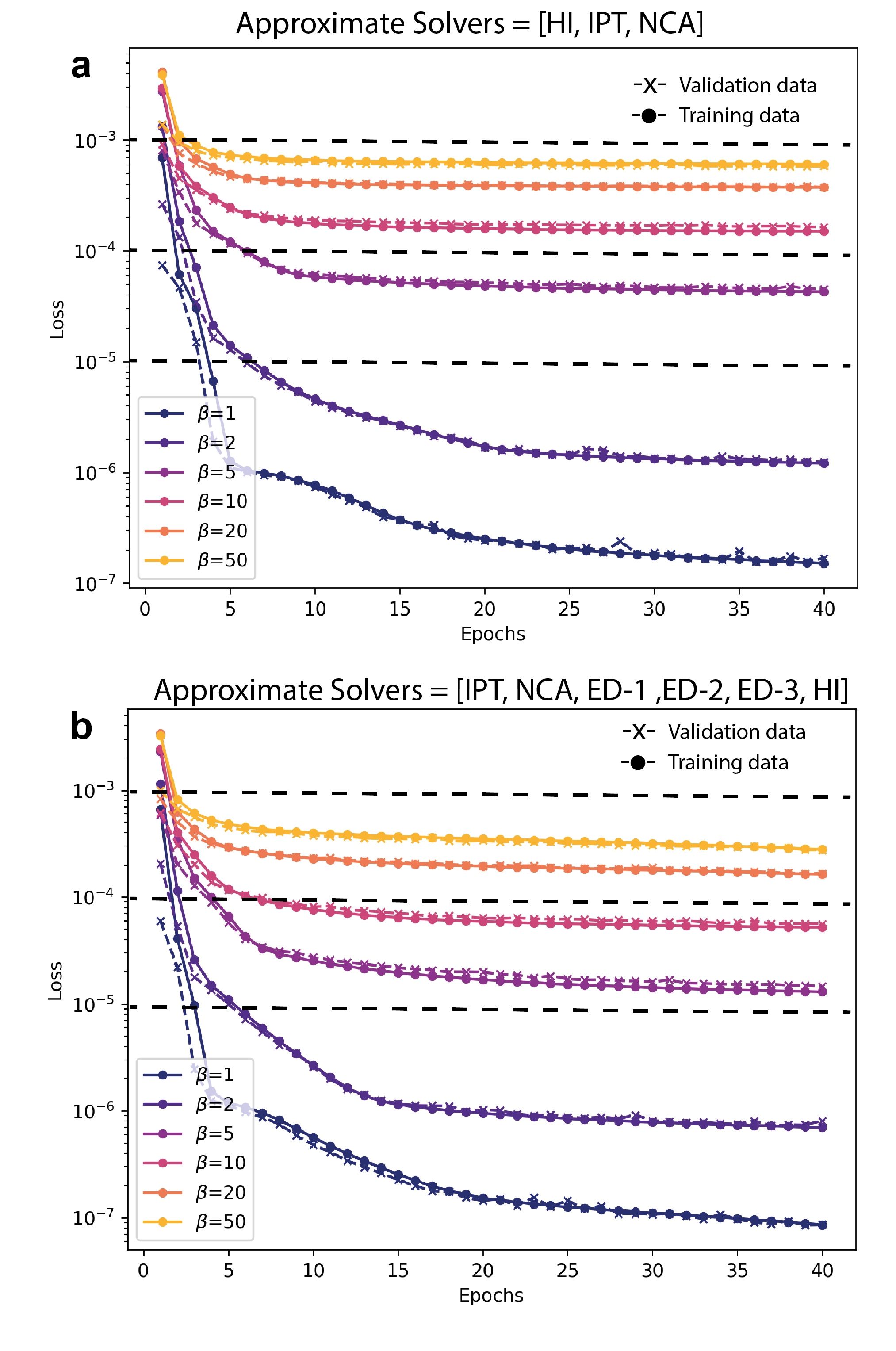}
  \caption{Cost function for the Legendre mesh for different inverse temperatures $\beta$ using as input to the neural network the solvers \textbf{a} [Hubbard-I, IPT, NCA] \textbf{b} [IPT, NCA, ED-1, ED-2, ED-3, Hubbard-I]. \edits{IPT is the Iterated Perturbation Theory Solver, NCA is the Non Crossing Approximation solver and ED-[1,2,3] are the truncated ED solvers.}}
  \label{fig:leg_loss}
\end{figure}

In Figure~(\ref{fig:leg_loss}) we show the value of the cost function when trained in the Legendre basis for models $\mathcal{M}_1$ and $\mathcal{M}_3$ using the $\mathcal{T}_1/2$ data transformation. We note that each value of $\beta$ necessitates a different mesh value $N_b$, as chosen according to the reconstruction criteria detailed in Figure~(\ref{fig:leg_coeffs}) of Appendix A. Moreover, the chosen mesh value $N_b$ is a critical factor in the complexity of the network, as determined by Eq~(\ref{eq:nn_num_params}). We observe similar results for the training in the Legendre basis, \emph{i.e} that higher temperatures are more amenable to the training procedure and that including more approximate solvers increases accuracy of the final validation loss. Notably, we see that the overall value of the loss function for $\mathcal{M}_3$ and $\mathcal{M}_1$ is lower than that predicted in the imaginary-time basis, regardless of the ensemble of solver types and transformation used.  Therefore, we see that by executing suitable basis transformations which are supplemented by a multitude of different approximate solvers then the accuracy of the overall predictive quality of the neural network can be improved.

In Appendix E, we describe the prediction of the optimised network for three
representative samples, respectively at small $U \approx 1.88-1.99$eV and larger $U\approx 6.2$eV in both the Legendre and imaginary-time bases at different temperatures. Of course, we have presented only two possibilities of AIM configurations, where
in principle the AIM parameter space is much larger. In general, while it is true
the predictive quality in the entire phase-space of $\{ U, W, \varepsilon \}$ is
poorer for larger values of $\beta$ (\emph{c.f}
Fig. \ref{fig:final_valid_loss}), we have demonstrated that it is still
better than what approximate solvers can achieve for some parameter
regimes, and even sometimes with effectively error-free solutions in both the imaginary-time and Legendre polynomial bases. Hence, \textbf{d$^3$MFT} provides a systematic improvement over the studied parameters and in many situations.

Moreover, it is noteworthy that the AIM parameter space for real materials is
quite restricted, where for example, due to the screened Coulomb interaction,
there are few materials with large Hubbard $U$ values, which lie in the  Mott limit, 
with a screened Coulomb interaction $\approx 8-10$eV. The archetypal La$_2$CuO$_4$ is 
one of the notorious case, where large $U$ effects are obtained by the strong 
localisation of the atomic d-shell. Most charge transfer correlated systems 
however have Coulomb interactions in the $3-6$eV range, poising \textbf{$d^3$MFT} as a 
good candidate for high throughput calculations in such cases. This same line of thinking applies for $\{ W, \varepsilon \}$ too, and opens up this approach for further dimensional reduction by taking only physically 
relevant parameter regimes into account.


\section*{III. Results: Data Driven Dynamical Mean Field Theory - d$^{3}$MFT }

\begin{figure}[h!] \centering
\includegraphics[width=\columnwidth]{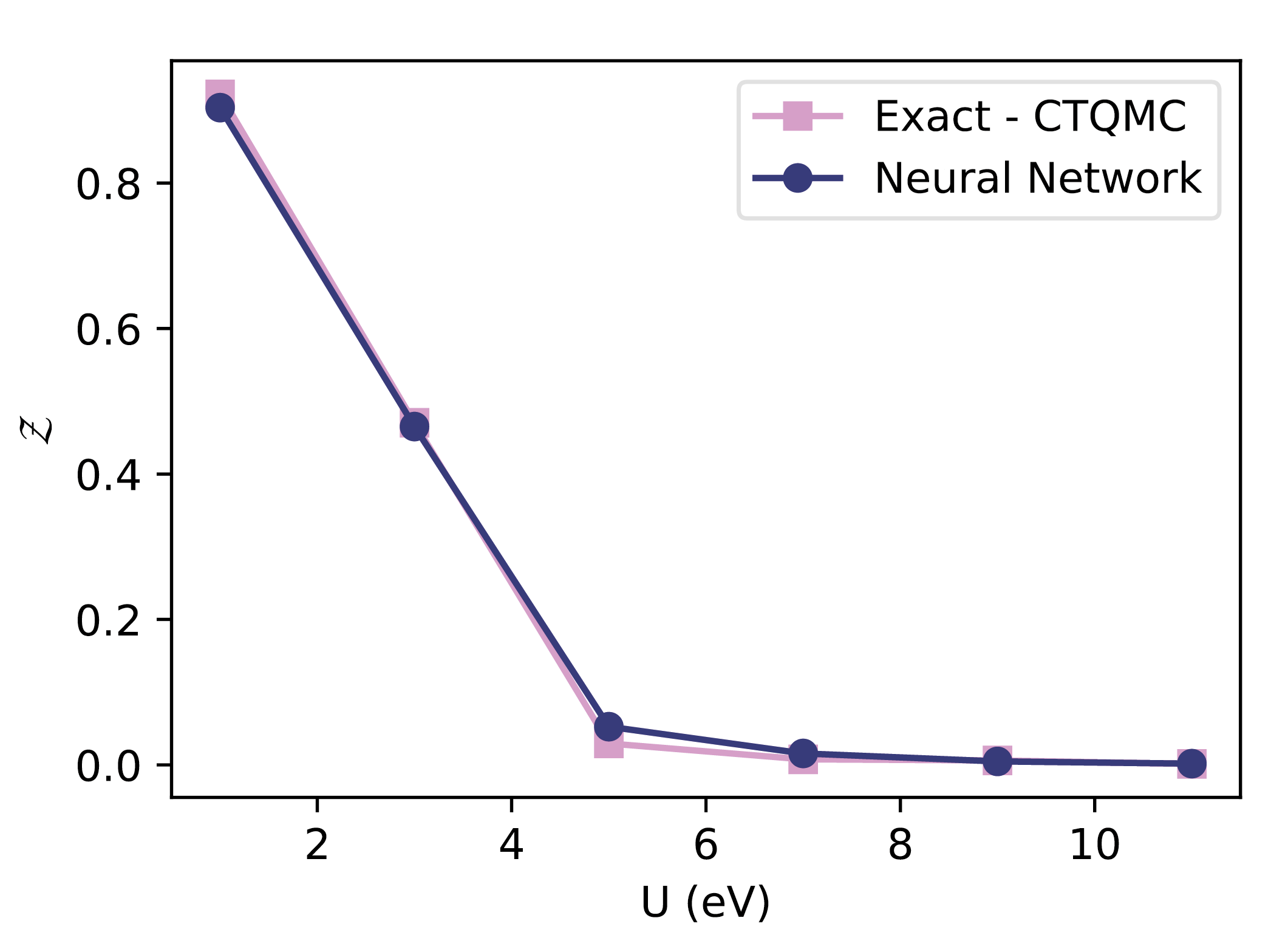}
  \caption[Capturing the Mott transition in the single-band Hubbard model using a Neural Network]{CTQMC and Neural Network solvers used for a DMFT prediction of the quasiparticle weight $\mathcal{Z}$ as a function of $U$ at $\beta=20$ eV$^{-1}$ and $W=1.0$ eV for the single-band half-filled Hubbard model on a Bethe lattice.}
  \label{fig:nn_mott}
\end{figure}

The motivation behind developing the data-driven impurity solver (\textbf{d$^{3}$MFT}) is to 
alleviate DMFT calculations from the intensive computational burden imposed by Exact 
Diagonalisation and Monte Carlo methods. In Fig. \ref{fig:nn_mott} we illustrate how the neural
network solver $\mathcal{M}_1$ in the imaginary time basis with an adapative mesh, used in a DMFT calculation, can predict the Mott transition at $\beta=20$ eV$^{-1}$, $W=1.0$ eV (at half-filling). This is compared with
the equivalent exact CTQMC results produced with the TRIQS/CTHYB solver \cite{triqs_cthyb}.
We choose to use the CTQMC solver as an independent verification since the neural 
network was trained on an ED solver. \edits{We use the quasiparticle weight to perform this comparison, defined by:}

\begin{equation}
  \edits{\mathcal{Z} = \frac{1}{\left(1 - \lim_{i \omega_n  \rightarrow 0}\frac{d\Im m \Sigma(i \omega_n)}{d i \omega_n}\right)}}
  \label{eq:QPW}
\end{equation}

\begin{figure}[h!] \centering
\includegraphics[width=\columnwidth]{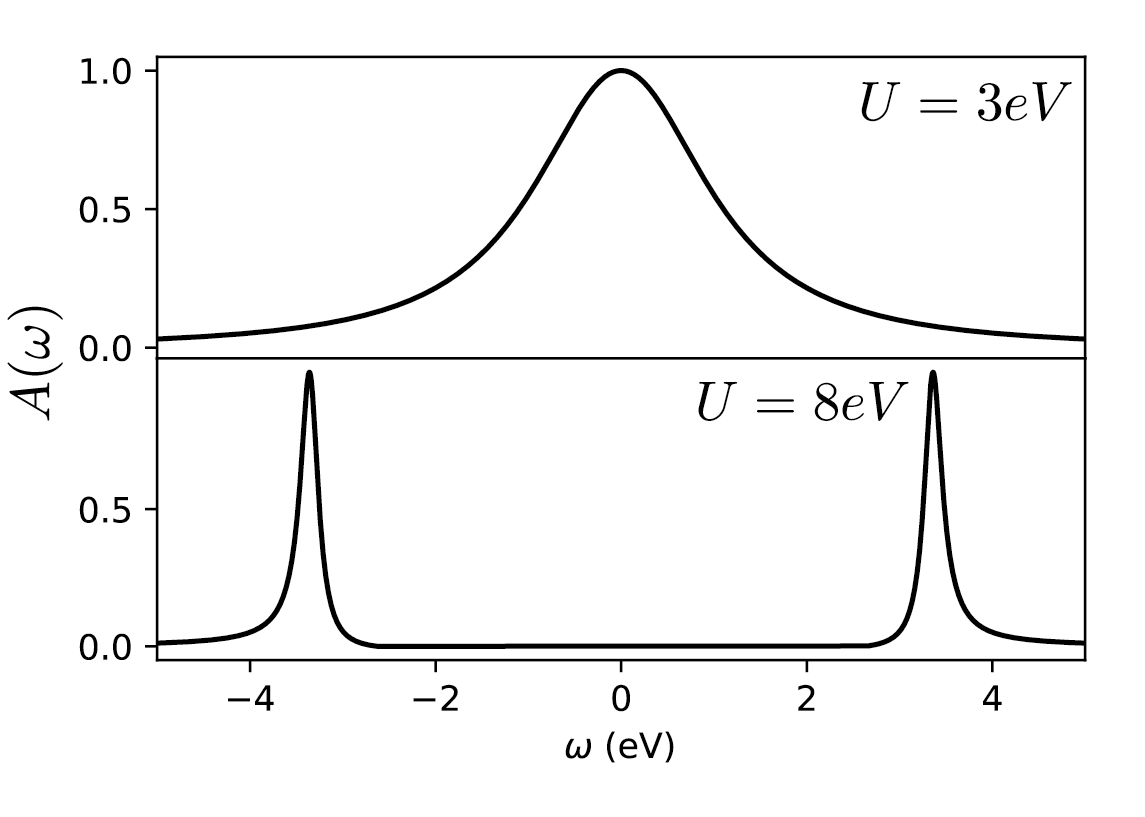}
  \caption{\edits{The resultant spectral functions as computed with the Neural Network solver for values of $U=3$ eV and $U = 8$ eV, $\beta=20$ eV$^{-1}$ and $W=1.0$ eV using the Pade method for analytical continuation. }}
  \label{fig:nnDOS}
\end{figure}

For each value of $U$, both solvers are run for $30$ iterations to a self-consistent solution. 
As $U$ is increased the Mott transition occurs at $U/D \approx$ \edits{$6$}, consistent with other calculations in the literature \cite{dmft_rev_mod}, \edits{up to a factor of 2 due to the choice of of $D=2eV$}. We emphasise that 
the network uses approximate solutions as its input during its cycle, for which 
it predicts the error-free corrected output instantly. By contrast the CTQMC has to be 
run long enough to remove its statistical error bars. \edits{We also present in Fig. \ref{fig:nnDOS}  the spectral functions obtained by analytic continuation performed on the neural network with the Pade method, as the neural network provides noise free solutions. The metal-insulator nature of the transition is well-captured with the strong signatures of Hubbard bands at U=8eV.} This proof-of-concept calculation
highlights the power of the data-driven method for a prototypical strongly correlated system. 

\section*{IV. Conclusion}

We have developed an extensible data-driven framework that can readily create error-free
neural network impurity solvers capable of capturing the Mott transition using DMFT for the Hubbard model. We call our new method \textbf{d$^{3}$MFT}.  Compared to standard machine learning methods our approach has a relatively small numerical footprint and is straightforward to implement,
once the set of impurity solvers are specified. As expected, and in line with
Ref. \cite{Arsenault_PRB_2014_ES}, the convergence and overall predictive quality
of the training procedure is sensitive to the basis representation of the
impurity Green's function. Using an adaptive imaginary-time $\tau$ mesh or Legendre coefficients under a \emph{tanh}-type basis transformation to express the Green's function, supplemented with data-augmentation techniques through exploiting physical symmetries, is the most effective route to minimising the overall error of the
network's cost function. Nevertheless, using the raw untransformed datasets in either 
of the two bases can still yield impressive predictive
capabilities, often in better agreement with the exact solution over the approximate solvers. The network shows a strong dependence on the
temperature at which the database is built for as well as the approximate solvers used. 
We find that training at higher temperatures using  approximate solutions 
results in consistently reliable results. We anticipate that improved 
results at higher temperatures could be attained by extending into the intermediate 
representation \cite{irbasis}, which is currently the most compact representation 
of the Green's function available. However, we also mention that since the 
entire protocol is a supervised learning method, it implicitly depends on 
the quality of the exact solution obtained. Therefore, the quality of the
data-driven solver produced at the end is limited by this and will always 
rely on the precision of the exact model that is used.

A natural extension to the single-band results presented here is to generalise
the method for a multi-orbital system. This will enable material-specific 
calculations to be undertaken rather than model Hamiltonians. However, the
extension to multi-orbital physics, especially for realistic materials, must take
into account various processes that arise from the Slater-Kanamori
representation, such as Hund's coupling. Furthermore, this aspect means that the
simplicity of the perturbative solutions is lost, which will have an effect on
correcting their errors in the training process. \edits{Additional extensions for this machine learning based approach includes a systematic study of doping, temperature and magnetism for the single-band Hubbard Model. To achieve this, the low-temperature prediction capabilities of the method need to be improved.} Notwithstanding these
considerations, we expect our proposed framework will be a valuable in
stimulating efforts in this direction, and ultimately complement the ongoing
research efforts to devise fast and accurate solvers for the AIM so that DMFT
calculations can be applied to a much broader class of problems that are out
of reach to current methods.

\quad

\section*{Acknowledgements}
CW acknowledges insightful and stimulating discussions with Andrew Mitchell.
CR thanks AWE for support through its Future Technologies fund. 
CW is supported by grant EP/R02992X/1  from the UK Engineering and Physical Sciences Research Council (EPSRC). ES is supported by the EPSRC Centre for Doctoral Training in Cross-Disciplinary Approaches to Non-Equilibrium Systems (CANES, EP/L015854/1).
IR and FJ acknowledge the support of the UK government department for Business, Energy and Industrial Strategy through the UK national quantum technologies programme (InnovateUK Industrial Strategy Challenge Fund (ISCF) QUANTIFI project). 
This work was performed using resources provided by the ARCHER UK National Supercomputing Service and 
the Cambridge Service for Data Driven Discovery (CSD3) operated by the University of Cambridge Research Computing Service (www.csd3.cam.ac.uk), provided by Dell EMC and Intel using Tier-2 funding from the Engineering and Physical Sciences Research Council (capital grant EP/P020259/1), and DiRAC funding from the Science and Technology Facilities Council (www.dirac.ac.uk). 
\quad
\section*{Additional Information}
Correspondence should be addressed to Evan Sheridan (sheridev@tcd.ie) and Cedric Weber (cedric.weber@kcl.ac.uk).

%




\subsection*{Appendix A: Data augmentation and data symmetrisation}
Once the database of approximate and exact solutions is constructed, and before
the data is passed to the machine learning algorithm for training, there are a
number transformation operations that allow the database to be augmented through
symmetry operations, without having to re-run the impurity solver. In this
appendix we explain a number of ways to both extend and transform the database
in ways that are amenable to learning a model. For simplicity, we assume a
database under consideration is expressed in either the imaginary-time basis on
a discrete, evenly spaced grid or the Legendre polyniomial basis.


The first symmetry operation makes use of the fact that the Green's function can be
decomposed into its symmetric and anti-symmetric contributions by decomposing it
into the Legendre basis, 
\begin{multline}
  G^{\text{S}}(\tau) = \sum_{\substack{l \geq 0 \\ \text{even}}}\frac{\sqrt{2l+1}}{\beta} P_{l}[x(\tau)]G_{l} \\ \text{ and } \\ G^{\text{AS}}(\tau) = \sum_{\substack{l \geq 0 \\ \text{odd}}}\frac{\sqrt{2l+1}}{\beta} P_{l}[x(\tau)]G_{l},
  \label{eq:legendre_rep_symm}
\end{multline}
where $G^{\text{S}}(\tau)$ and $G^{\text{AS}}(\tau)$ are respectively the
symmetric and anti-symmetric parts of total Green's function which give the total
Green's function when summed \emph{i.e.},
\begin{equation}
  G(\tau) = G^{\text{S}}(\tau) + G^{\text{AS}}(\tau).
  \label{eq:gtau_symm_antisymm}
\end{equation}

\begin{figure}[h!]
  \centering
  \includegraphics[width=\columnwidth]{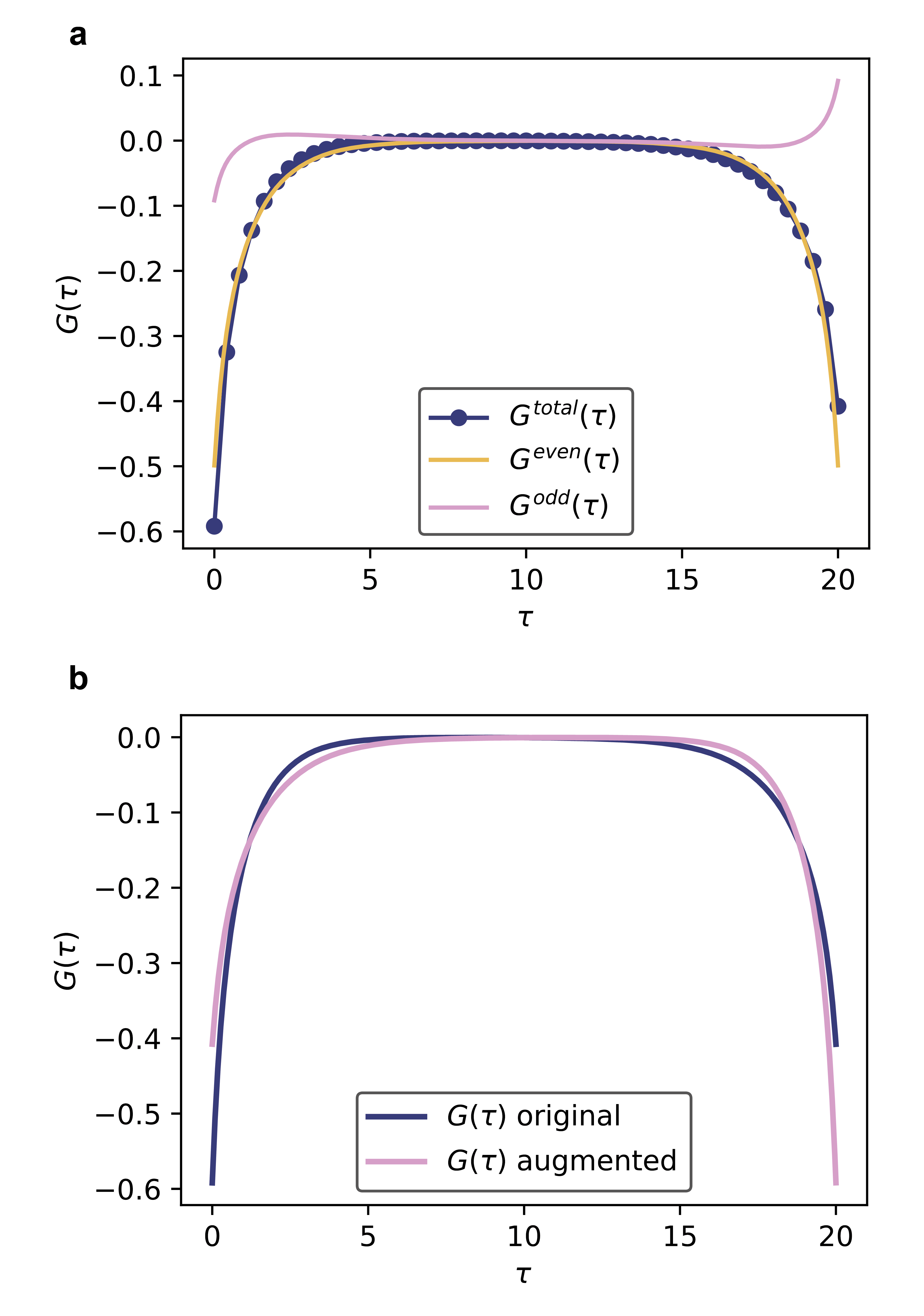}
  \caption[Data augmentation schemes for $G(\tau)$]{Data augmentation schemes for $G(\tau)$ where \textbf{a} illustrates the extraction of the odd and even parts of $G(\tau)$ via Legendre transformations and \textbf{b} exploiting the particle-hole symmetry of $G^{e}(\tau)=G^{h}(\beta-\tau)$ where $G^{e}(\tau)$ is the electron Green's function and $G^{h}(\tau)$ is the hole Green's function}
  \label{fig:data_augmentation}
\end{figure}

In practice, if performed in the $\tau$ basis this requires an intermediate step
of generating the Legendre coefficients, or reading them in from a database which
has them stored already. For the size of the databases dealt with in this
study ($<40,000$) this feature can be added practically at no additional
computational cost. Physically, the symmetric part of the Green's function
represents the physics at half-filling while the anti-symmetric component
encodes the information away from half-filling. This operation need not only be
used for the augmentation of the database, it can similarly be used for
partitioning it. Specifically, instead of training a model on both the symmetric
and anti-symmetric components simultaneously, two separate models can be
trained on the symmetric and anti-symmetric components separately, after
which they are recombined to produce the total answer in Eq. 
(\ref{eq:gtau_symm_antisymm}). The same procedure can be followed in the
Legendre basis, where the symmetric part of $G_l$ is encoded in the even
coefficients and the anti-symmetric part in its odd ones. For both bases, 
this operation allows the database to be augmented by a factor of two and is
shown in Fig. \ref{fig:data_augmentation}a.

The second symmetry operation on $G(\tau)$ that we can consider is particle-hole
equivalency, \emph{i.e.} $G^{e}(\tau)=G^{h}(\beta-\tau)$ where $G^{e}$ is the
electron Green's function and $G^{h}$ is the hole Green's function. Therefore,
for every entry in the database that is away from half-filling, the
corresponding electron (if hole-type) or hole (if electron-type) Green's
function can be generated simply by flipping $G(\tau)$, and is illustrated in
Fig. \ref{fig:data_augmentation}b. If, however, the database is expressed in
the Legendre basis instead, this transformation requires that the odd
coefficients be multiplied by $-1$. We note that for both basis 
representations, this can double the size of the database. 

\subsection*{Data transformation}

In addition to augmenting the database by exploiting symmetry operations, the
data must also be transformed into a representation appropriate for how the
training data will be manipulated. Here we outline the common ways in which the
data are transformed. For lower temperatures the shape of $G(\tau)$ becomes
steeper at its boundaries and flatter further away. If a regularly spaced out
$\tau$-mesh is used this presents a problem as the behaviour near the boundary
may not be taken into account, or in the worst case missed completely. From a
data-centric perspective, this would mean that different $G(\tau)$ functions
which describe qualitatively different physics could look very similar to a
learning model, since most of the data passed to it will be constant. To remedy
this issue it is possible to transform the $\tau$ domain so that it is sampled
more near its boundaries and less near the centre. This can be achieved by
transforming the $\tau$ points from being evenly distributed to being
distributed according to a sigmoid-type function $\sigma$.

\begin{figure}[h!]
  \centering
  \includegraphics[width=\columnwidth]{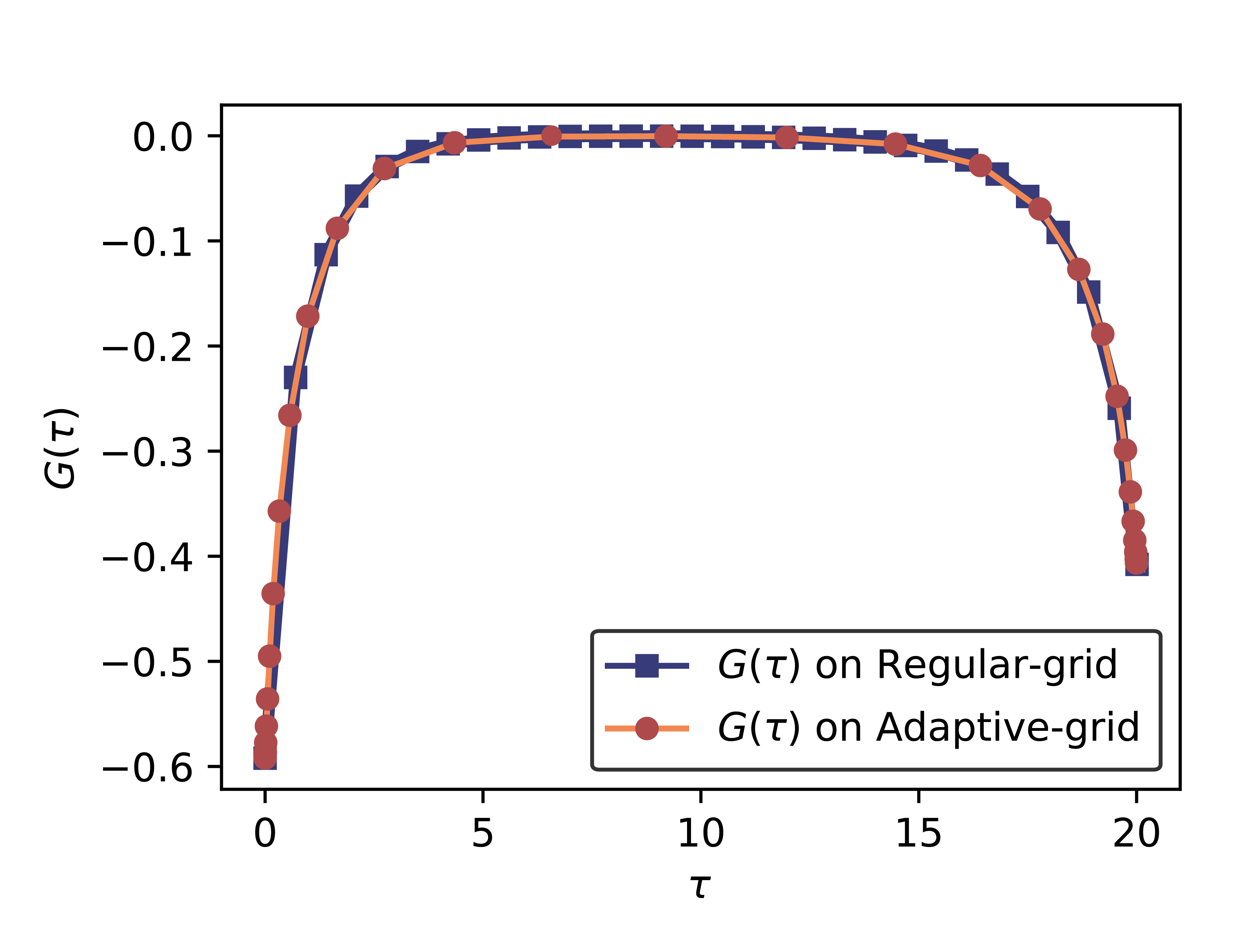}
  \caption[Adaptive $\tau$ mesh transformation]{Transformation of $G(\tau)$ from a regular equidistant mesh to an adaptive mesh.}
  \label{fig:adaptive_grid}
\end{figure}

Thus, the transformation from a regular equidistant mesh to an adaptive mesh
amounts to $G(\tau) \rightarrow G(\sigma(\tau,\delta))$, where $\delta$ is a smearing parameter,  and is illustrated in
Fig. \ref{fig:adaptive_grid} for a typical $G(\tau)$ insulating type
solution of the AIM. In this case, if a regular mesh is used then more than
$50\%$ of all data-points are zero, due to $G(\tau) \approx 0$ between
$5<\tau<15$. However, if the adaptive mesh is used less then $10\%$ of the
points fall in this interval are zero. Indeed, depending on the inverse
temperature $\beta$, the optimal choice of $\delta$ should ensure that the
features of $G(\tau)$ are retained away from where it plateaus. In practice, the
procedure to obtain $G(\sigma(\tau,\delta))$ is as follows:

\begin{enumerate}
  \item Spline the equidistant $G(\tau_1,\hdots,\tau_N)$ to get its continuous representation. 
  \item Define the new adaptive grid $\sigma(\tau, \delta)$ from $[0, \beta]$ on a continuous grid.
  \item Evaluate $G(\sigma(\tau,\delta))$ on the continuous grid.
  \item Sample $G(\sigma(\tau,\delta))$ $N$ times to obtain $G(\tau_1^{\text{ad}}, \hdots, \tau_N^{\text{ad}})$, where  $\tau_{i}^{\text{ad}}$ indicates the $i$'th $\tau$ point on the adaptive mesh. 
\end{enumerate}

\begin{figure}[h!]
  \centering
  \includegraphics[width=\columnwidth]{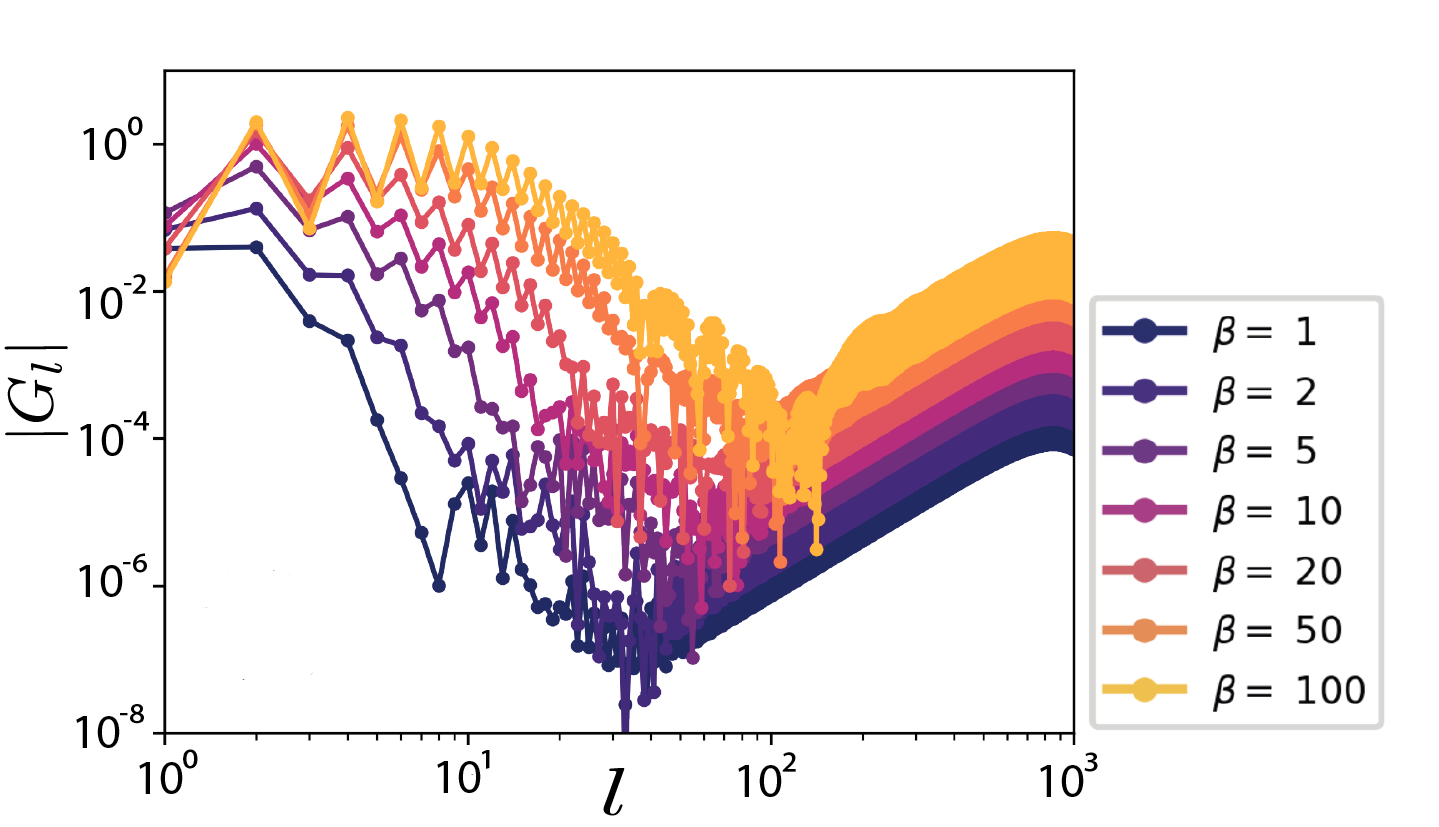}
  \caption[Legendre coefficients as a function of different inverse temperatures]{The absolute value of the Legendre coefficients $|G_l|$ used to express the Green's function of the Anderson Impurity Model at a range of different inverse temperature values $\beta$ on a log-log scale.}
  \label{fig:leg_coeffs}
\end{figure}

We note that adaptive meshes are not required if the Green's function is
expressed in a polynomial basis. There is still, however, the issue of obtaining
spurious Legendre coefficients if a cutoff of $n_{l}^{\text{max}}$ is not imposed. We
illustrate this problem in Fig. \ref{fig:leg_coeffs} for a range of
temperatures using $10^{3}$ Legendre coefficients for each one. For $n_l>10^{2}$
it is clear that including more coefficients in the expression for $G(\tau)$
starts to exhibit classical over-fitting behaviour. Therefore, when using the
Legendre basis a judicious choice of $n_{l}^{\text{max}}$ must be made,
otherwise the model will be trained on data that lacks relevant physical information. As depicted in
Fig. \ref{fig:leg_coeffs} when $\beta$ increases so do they number of
Legendre coefficients needed for an accurate expression of the Green's function.
Therefore, depending on the temperature chosen, the size of the feature space
changes for the Legendre representation. Whereas in the case of the
imaginary-time basis, as $\beta$ increases the necessity to use an adaptive mesh
becomes higher.

Finally, we now turn to the final set of data transformations that must be
performed. Scaling the input and output variables so that they are normalised is
a standard technique when preparing data for training algorithms such as 
neural networks. The reason for this is to protect the weights that are learned
in the model from becoming too large or biased towards large input values.
Specifically, this is essential for when input variables are the Legendre
coefficients, as the Legendre basis has no inherent scale for the coefficients.
On the other hand, while an inherent scale exists for $G(\tau)$, \emph{i.e.} $ -1
\leq G(\tau) \leq 0$ when $\tau>0$, it is also possible to create a family of
scaling transformations and test their efficacy throughout the training process.
The following scaling transformations work regardless whether the aforementioned
symmetrisation or augmentation procedures have been followed.

\begin{figure}[h!]
  \centering
  \includegraphics[width=\columnwidth]{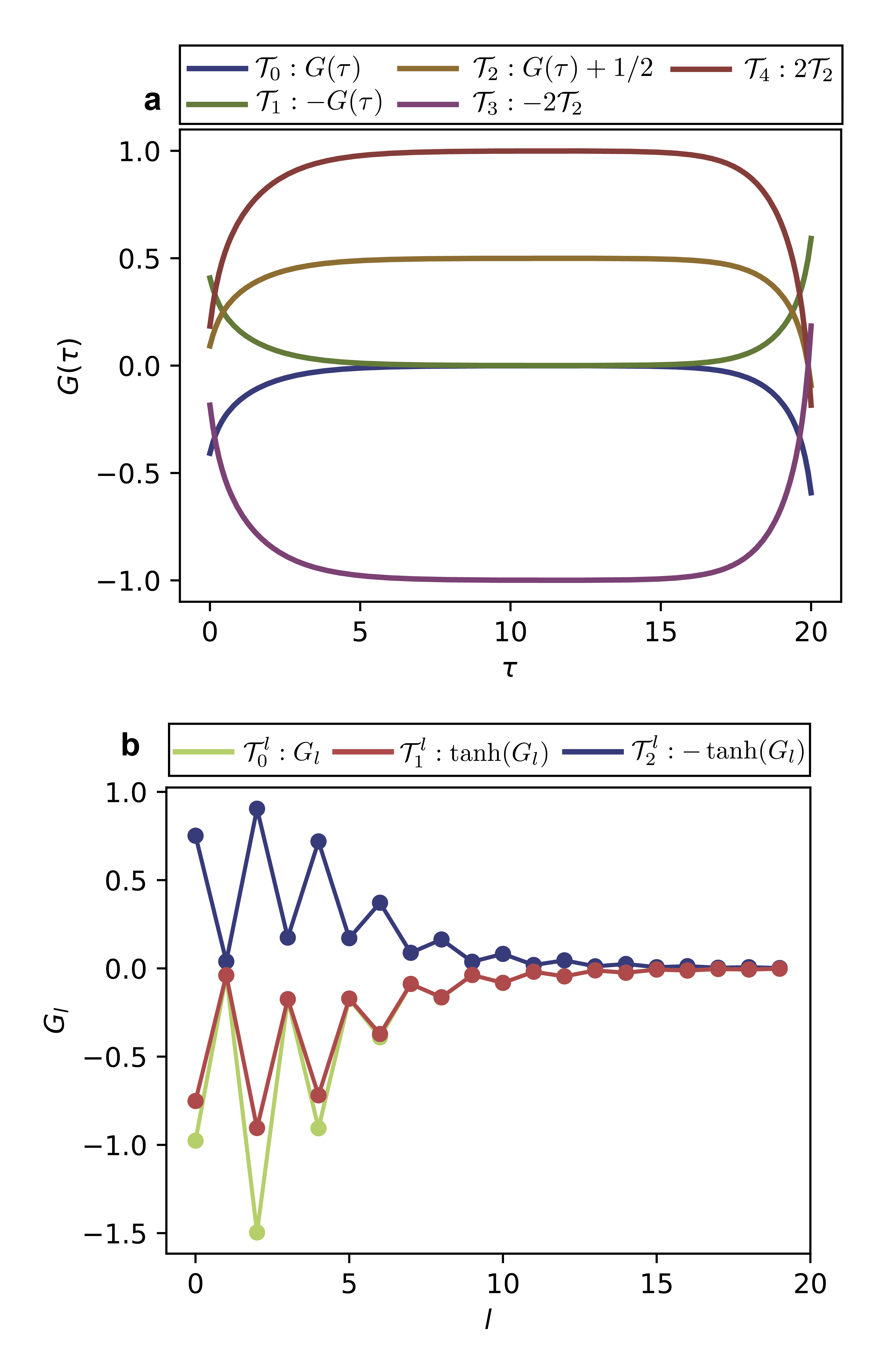}
  \caption[Table of scaling transformations for $G(\tau)$ and $G_l$]{Different scaling transformations before the data is trained on for the \textbf{a} imaginary-time Green's function $G(\tau)$ and \textbf{b} Legendre Green's function $G_l$}
  \label{fig:data_splitting}
\end{figure}

\begin{figure}[h!]
  \centering
  \includegraphics[width=\columnwidth]{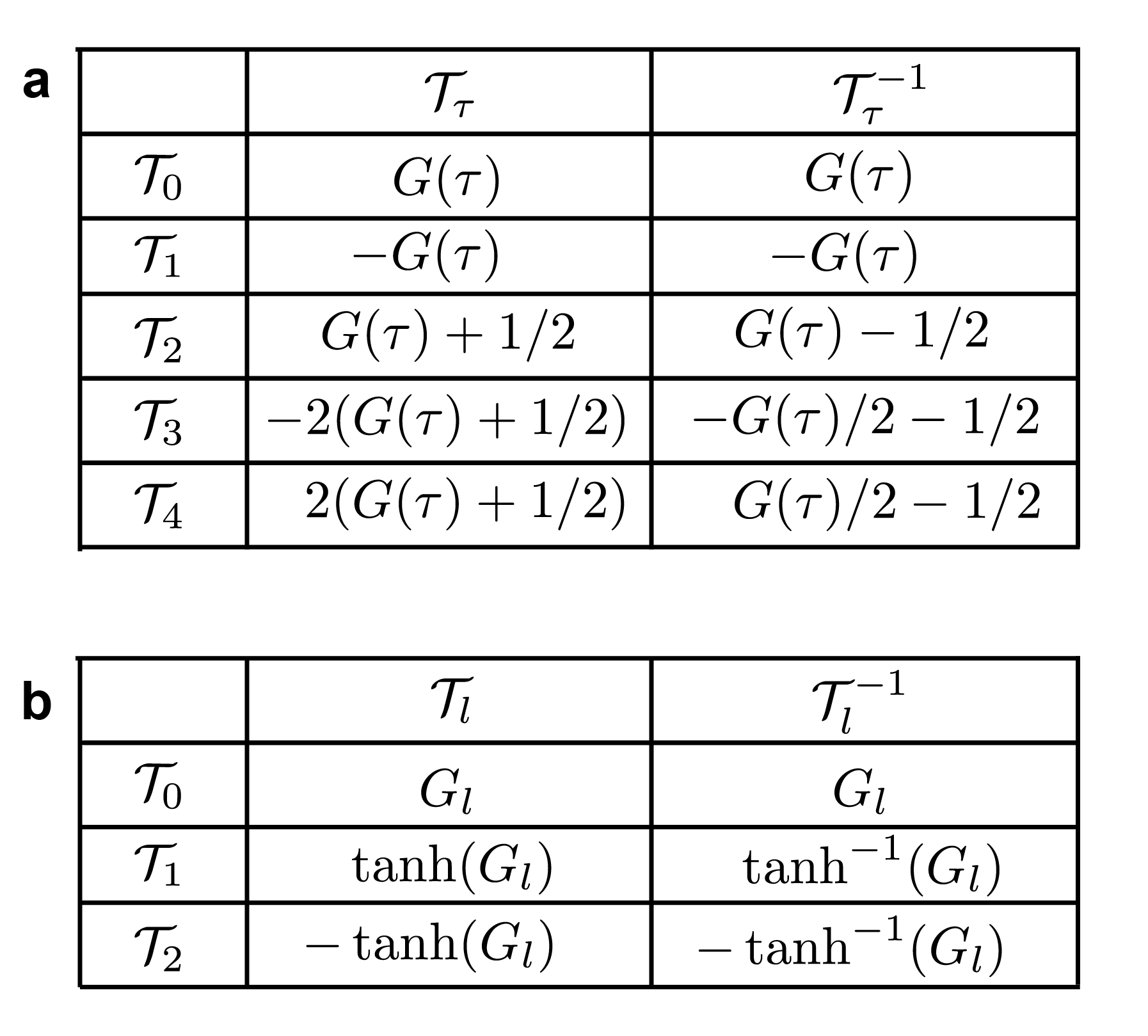}
  \caption[Depiction of scaling transformations for $G(\tau)$ and $G_l$]{Scaling transformations for \textbf{a} $G(\tau)$ and \textbf{b} $G_l$}
  \label{fig:trans_table}
\end{figure}

We first discuss the transformations $\mathcal{T}$ for $G(\tau)$ as 
shown in Fig. \ref{fig:data_splitting}a. $\mathcal{T}_0$ is the
unscaled Green's function and each transformation is a function of
$\mathcal{T}_0$. For $G(\tau)$ the situation is quite simple, there are only a
few transformations that can be done to normalise in between the range $[0,1]$
or $[-1,1]$. We note that if $G^{\text{AS}}(\tau)$ is used, \emph{i.e.} the
anti-symmetric part of the Green's function, it is important to ensure that the
scaling operations do not shift the data out of the scaling range, and so an
extra constant shift should be applied in these cases to counteract this
behaviour. The Legendre basis and its scaling transformations are
illustrated in Fig. \ref{fig:data_splitting}b. In this case it is clear that the
unscaled data is not normalised. Fortunately, by applying a \emph{tanh} function
this can readily be achieved. In the example shown, we see the
first anti-symmetric component of the Green's function, $G_1$, is scaled to be
much closer to $G_0$ and $G_2$, the either-side symmetric components. As stated
above for the $\tau$ basis, the dependence of training the model is also
assessed as a function of these transformations. Moreover, we emphasise that to
recover the physical Green's function it is necessary to apply the relevant
inverse transformation $\mathcal{T}^{-1}$, which are tabulated in
Fig. \ref{fig:trans_table}.

\subsection*{Appendix B: error scaling}

\begin{figure*}[h!]
  \centering
  \includegraphics[width=1.8\columnwidth]{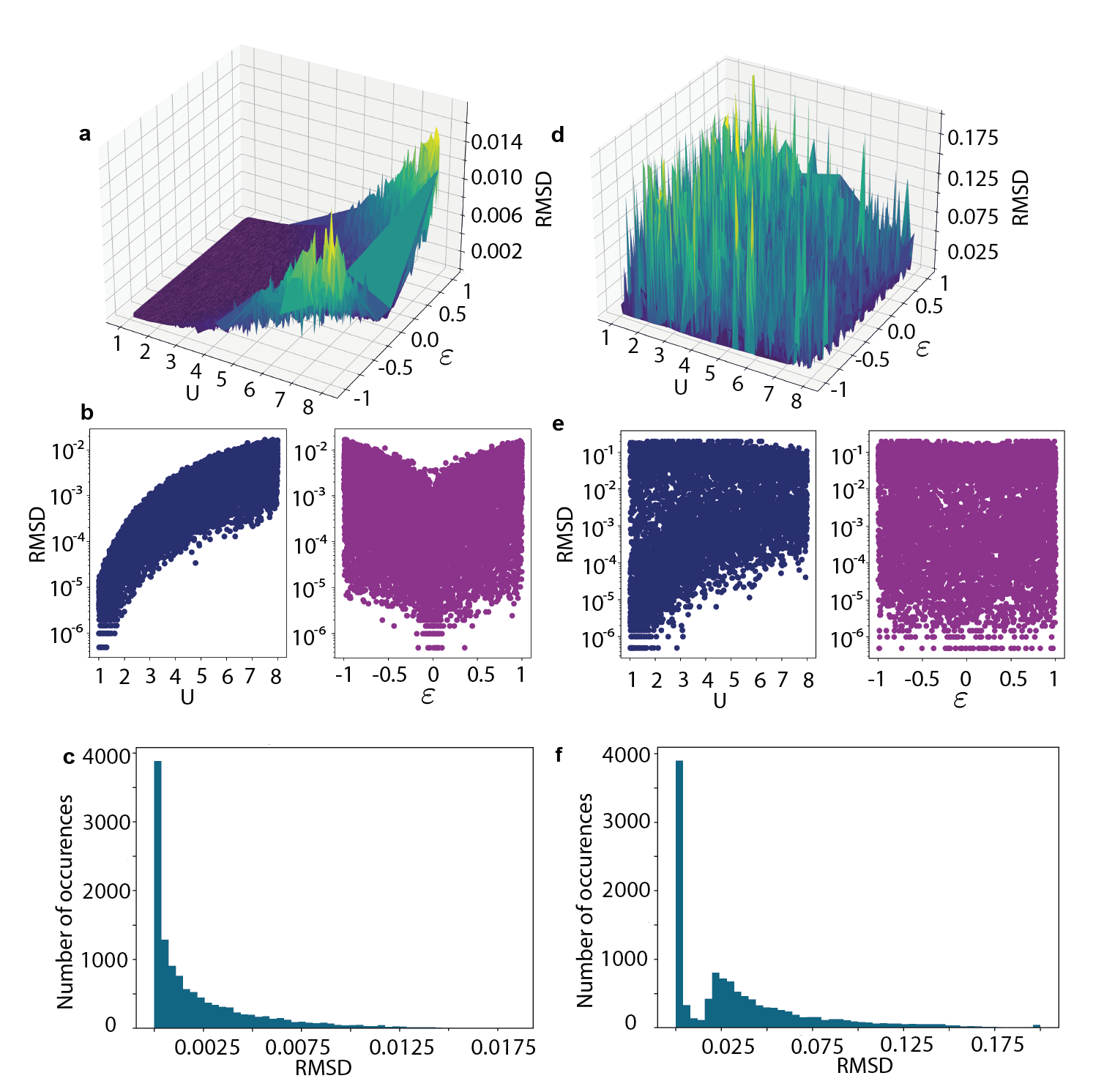}
  \caption[$U-\varepsilon$ error for the Iterated Perturbation Theory solver]{The $U-\varepsilon$ error of the impurity Green's function $G(\tau)$ calculated with the Iterated Perturbation Theory solver relative to the Exact Diagonalisation solution. For $\beta=1$ eV$^{-1}$ \textbf{a} 3d-cut of the error \textbf{b} 2d-cut of the error \textbf{c} histogram of the error. For $\beta=20$ eV$^{-1}$ for \textbf{d} 3d-cut of the error \textbf{e} 2d-cut of the error \textbf{f} histogram of the error }
  \label{fig:error_impurity_solver}
\end{figure*}

As discussed in this work, the error shows a dependence on the chosen
parameters $\{ U, \varepsilon \}$. Here we investigate the overall
features of this dependence by examining it over the entire phase space, restricting the U-values of up to 8eV that represent physical realisable scenarios. To this end, we explore the $U-\varepsilon$ phase-space of the error, chosen to be
the RMSD, between the approximate solvers and exact one. We select
$U-\varepsilon$ as this combination characterises the physics of the impurity,
which is what is being perturbed for the Hubbard-I, IPT and NCA solvers, and hence is where
the error should manifest itself. In Fig.  \ref{fig:error_impurity_solver} we
explore this space for the IPT solver for $\beta=1$ eV$^{-1}$ and $\beta=20$ eV$^{-1}$. 
For $\beta=1$ eV$^{-1}$, illustrated in Fig.  \ref{fig:error_impurity_solver}a-c, we
see a clear structure of the error, which is $<10^{-5}$ for $U<4$ eV, and also at
near half-filling, clearly highlighted for the 2d-cut in panel \textbf{b}. For
large values of $U$, away from half-filling, the solver performance starts to deteriorate.
Indeed, this is the expected behaviour of the IPT solver\cite{dmft_rev_mod} for
these parameter regimes, and we see that there is a high density of points
falling near zero in panel \textbf{c} in its error.

On the other hand, for $\beta=20$ eV$^{-1}$ (Fig. \ref{fig:error_impurity_solver}d-e) a 
lot of the structure present at $\beta=1$ eV$^{-1}$ disappears. In particular, at 
low $U$ the error extends to larger RMSD values and the structure in $\varepsilon$ 
is practically lost. Interestingly, the RMSD error in panel \textbf{f} splits 
into two regions, one where the error persists to be low while the other is 
high, and highlights that lowering temperature amplifies the error up to an 
order of magnitude or greater. While there is still a persistent structure in 
the error as a function of $U$ at $\beta=20$ eV$^{-1}$, our results suggest that 
predicting it reliably will be more difficult.

\begin{figure*}[h!]
  \centering
  \includegraphics[width=1.8\columnwidth]{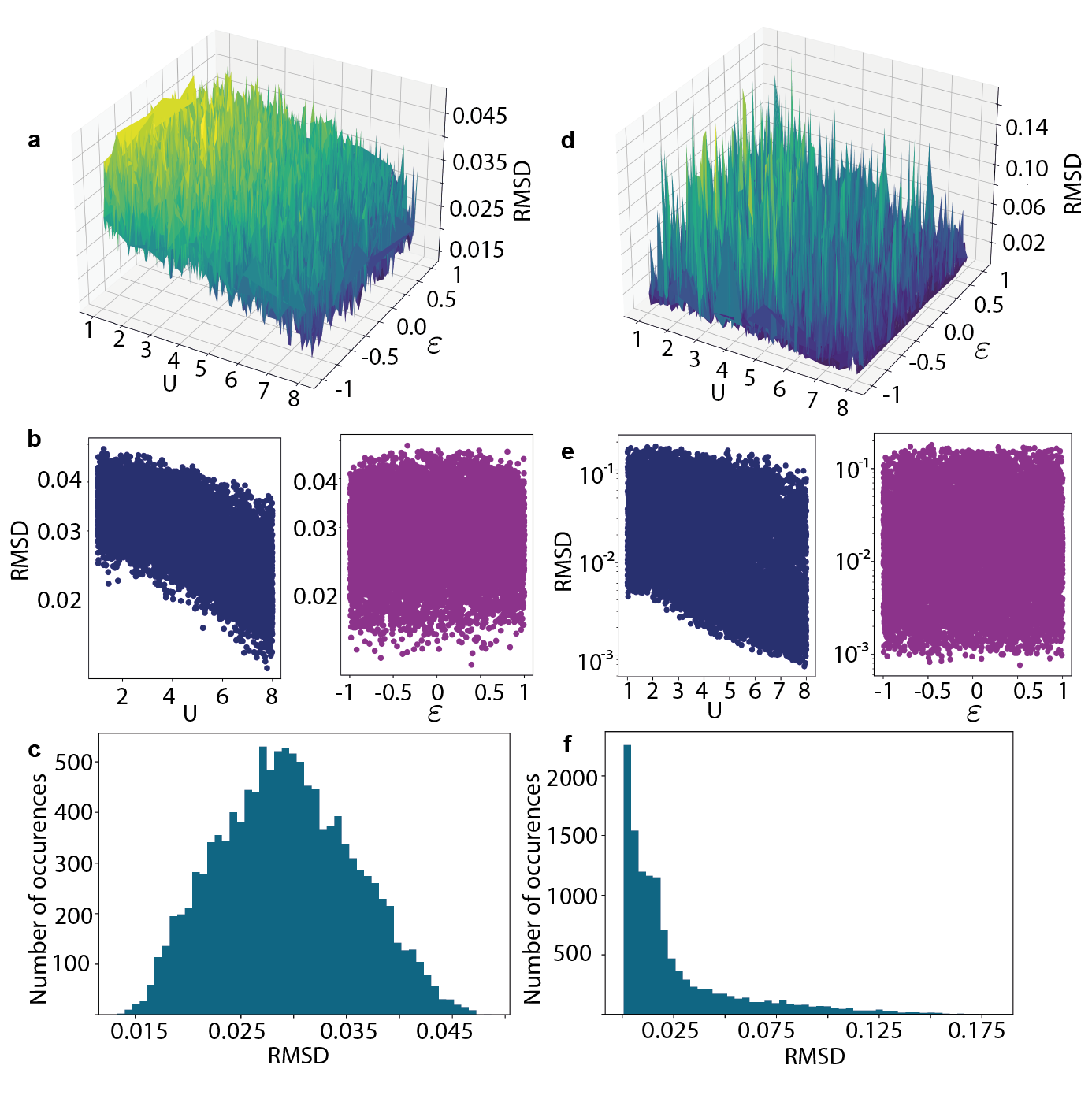}
  \caption[$U-\varepsilon$ error for the Non Crossing Approximation solver]{The $U-\varepsilon$ error of the impurity Green's function $G(\tau)$ calculated with the Non Crossing Approximation solver relative to the Exact Diagonalisation solution. For $\beta=1$ eV$^{-1}$ \textbf{a} 3d-cut of the error \textbf{b} 2d-cut of the error \textbf{c} histogram of the error. For $\beta=20$ eV$^{-1}$ for \textbf{d} 3d-cut of the error \textbf{e} 2d-cut of the error \textbf{f} histogram of the error }
  \label{fig:error_impurity_solver_NCA}
\end{figure*}

We repeat the same method of analysis for the NCA solver shown in Fig. 
\ref{fig:error_impurity_solver_NCA}. Again, for $\beta=1$ eV$^{-1}$ there are
systematic features to the error where it is lower for larger values of $U$, as
shown in Fig. \ref{fig:error_impurity_solver_NCA}a-b. Indeed, this is as
expected when using the NCA solver as it is a strong-coupling
expansion\cite{dmft_rev_mod}. Notably, even though the trend favours higher
$U$ values, their overall magnitude is larger than for the IPT solver by an
order of magnitude, even though the trend is not present there. We also notice
that the distribution of the error (panel \textbf{c}) is markedly different,
taking a Gaussian-like shape centred around the RMSD value of $0.03$. This
Gaussian-like behaviour is, however, absent in the $\beta=20$ eV$^{-1}$ case (panel
\textbf{f}) where the distributed error reverts back to decaying away
from zero. Moreover, in a similar fashion to the IPT case, for $\beta=20$ eV$^{-1}$ a
significant portion of structure disappears in the $U-\varepsilon$ dependence of
the error, which suggests that correcting for it may become increasingly difficult 
for lower temperatures.
 
\begin{figure*}[h!]
  \centering
  \includegraphics[width=1.8\columnwidth]{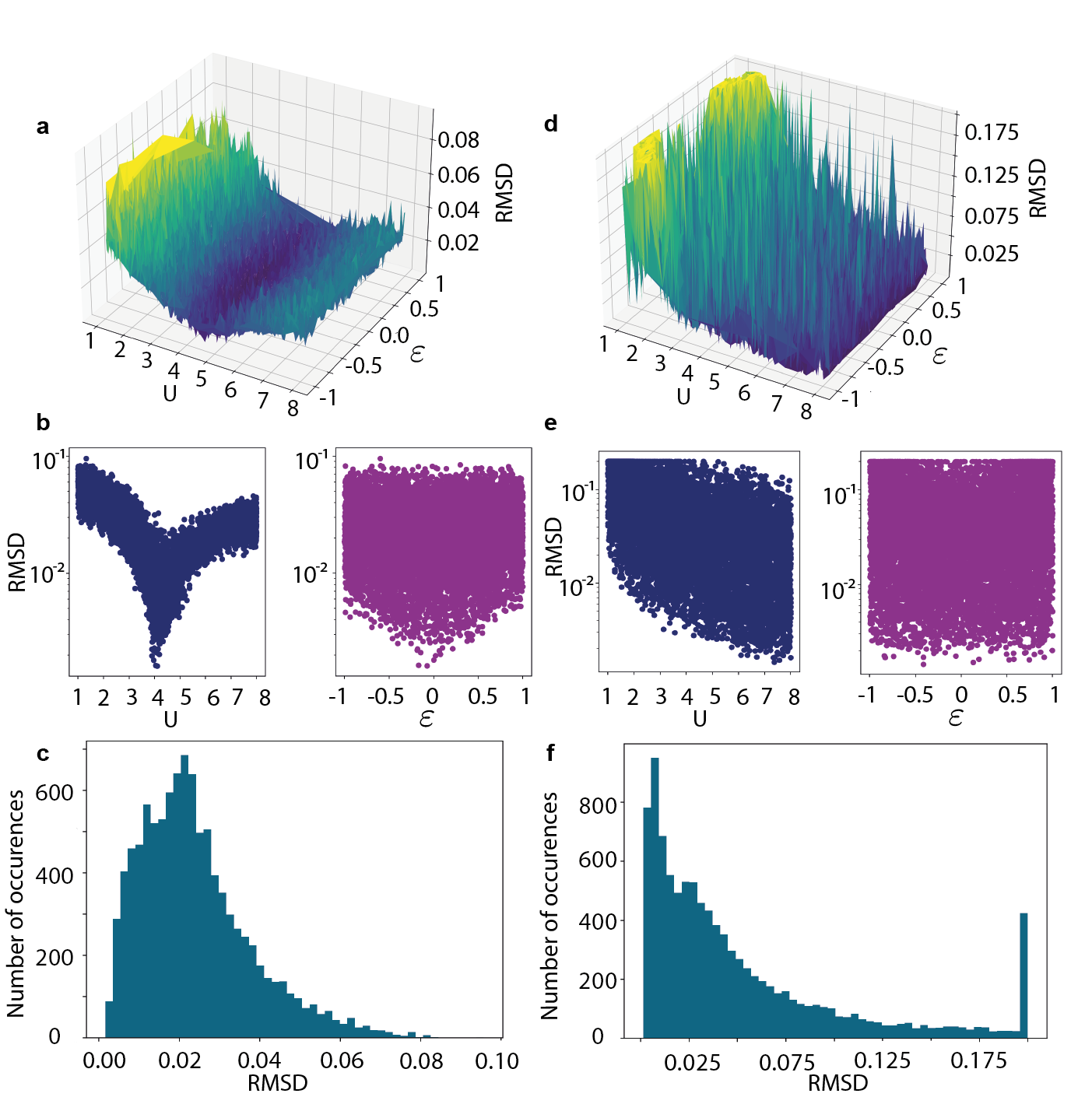}
  \caption[$U-\varepsilon$ error for the Hubbard-I]{The $U-\varepsilon$ error of the impurity Green's function $G(\tau)$ calculated with the Hubbard-I solver relative to the Exact Diagonalisation solution. For $\beta=1$ eV$^{-1}$ \textbf{a} 3d-cut of the error \textbf{b} 2d-cut of the error \textbf{c} histogram of the error. For $\beta=20$ eV$^{-1}$ for \textbf{d} 3d-cut of the error \textbf{e} 2d-cut of the error \textbf{f} histogram of the error }
  \label{fig:error_impurity_solver_Hi}
\end{figure*}

We repeat the same method of analysis for the Hubbard-I solver shown in Fig. 
\ref{fig:error_impurity_solver_Hi}. Similarly for the IPT and NCA solvers there is a noticable structure present at $\beta=1$ eV$^{-1}$, as shown in Figures~(\ref{fig:error_impurity_solver_Hi}a-c). Interestingly, we see that the error is minimal for $U\approx5$eV, which is opposite to what was observed for the IPT and NCA solvers. Moreover, at  $\beta=1$ eV$^{-1}$ the distribution of the RMSD error is heavy-tailed and shows a sensitivity away from half filling, similar to the IPT result, for example. Nevertheless, its magnitude is roughly about $10^{-2}$, more in line with the NCA prediction. Therefore, the Hubbard-I solution bears the hallmarks of capturing qualitative features of the exact solution at the intermediate interaction strengths, and places itself as a complementary approximate solver to the IPT and NCA solvers. Finally, we see that for $\beta=20$ eV$^{-1}$ the structure in the error deteriorates, but nevertheless retains some overall features in the error landscape, as seen in Figure~(\ref{fig:error_impurity_solver_Hi}d-f). Here, we see a striking resemblance to the NCA solver, whereby there is a noticeable structure in the error for large and small values of the Hubbard $U$ parameter.  

\begin{figure*}[h!]
  \centering
  \includegraphics[width=1.8\columnwidth]{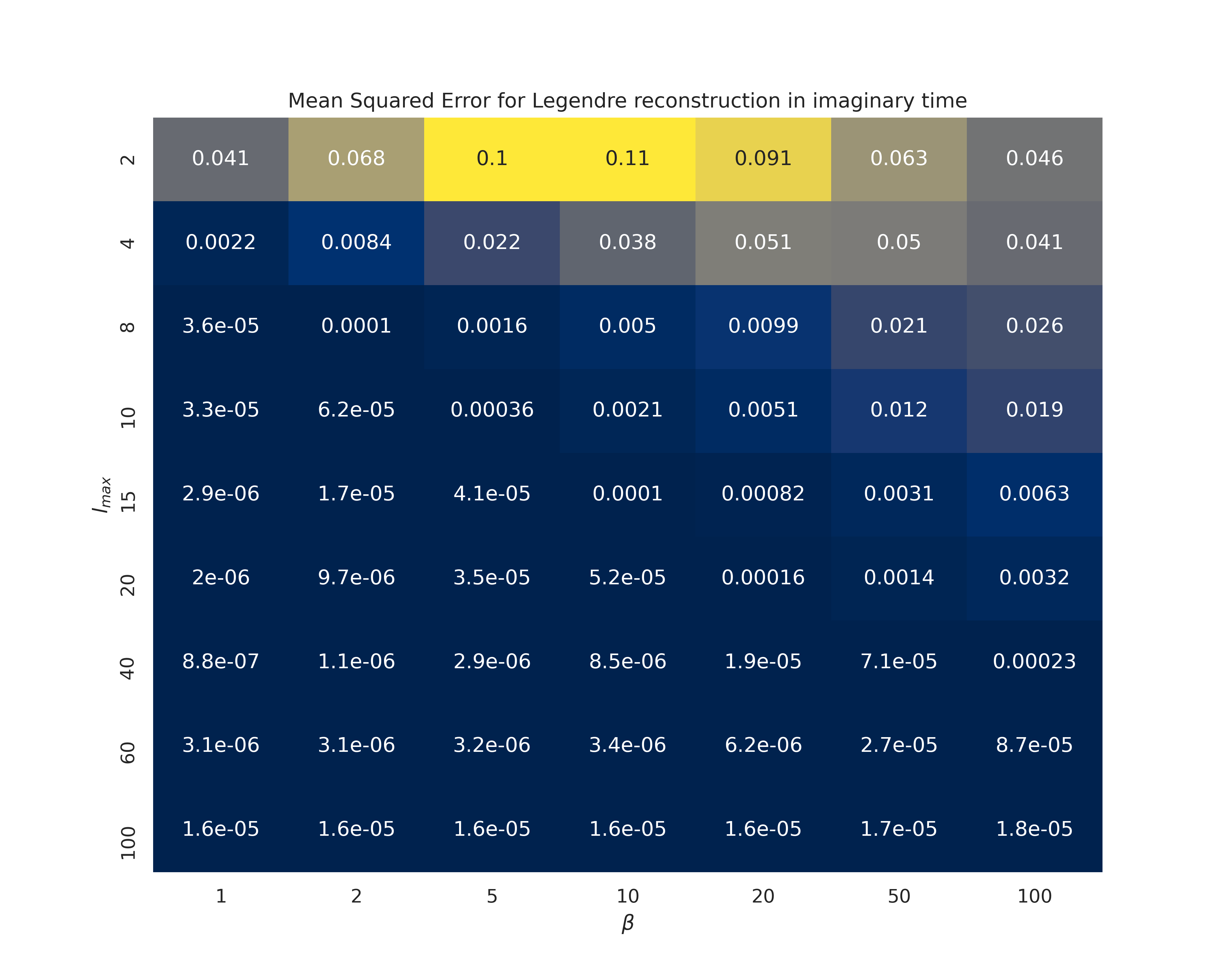}
  \caption[Error of reconstructed $G(\tau)$ using $l_{max}$ Legendre polynomials]{Reconstructed $G(\tau)$ from $l_{max}$ Legendre polynomials for the AIM heatmap as a function of inverse temperature $\beta$ and $l_{max}$. $l>60$ ensures for $\beta<100$ a completly accurate representation of $G(\tau)$.}
  \label{fig:legendre_heatmap}
\end{figure*}

We also note that the behaviour in the Legendre basis is strikingly similar,
although the $U-\varepsilon$ is far more discrete in regions where the solvers
perform well. Nevertheless, it is conclusive that a
basis-switch is not sufficient to recover the discernible features of its structure
at lower temperatures. Moreover, we clarify that the
$G(\tau)$ and $G_l$ data analysed is the raw data, without any 
processing, except for the truncation of the Legendre basis. We use Fig. 
\ref{fig:legendre_heatmap} to determine the $\beta$ dependence of the
this cut-off, where the input AIM is a sample from the generated database. In
order to maintain an error $<10^{-5}$ as function of increasing $\beta$ we
choose $l_{max}=8$ for $\beta=1$ and follow its diagonal along the matrix in Fig. 
\ref{fig:legendre_heatmap}. The over-fitting of $G(\tau)$ with $G_l$
is also evident here, where for example if $\beta=10$ as $l_{max}$ is increased
there is an eventual increase in the error for $l_{max}=100$. 

\subsection*{Appendix C : Quantum database samples}

\begin{figure*}
\includegraphics[width=1.5\columnwidth]{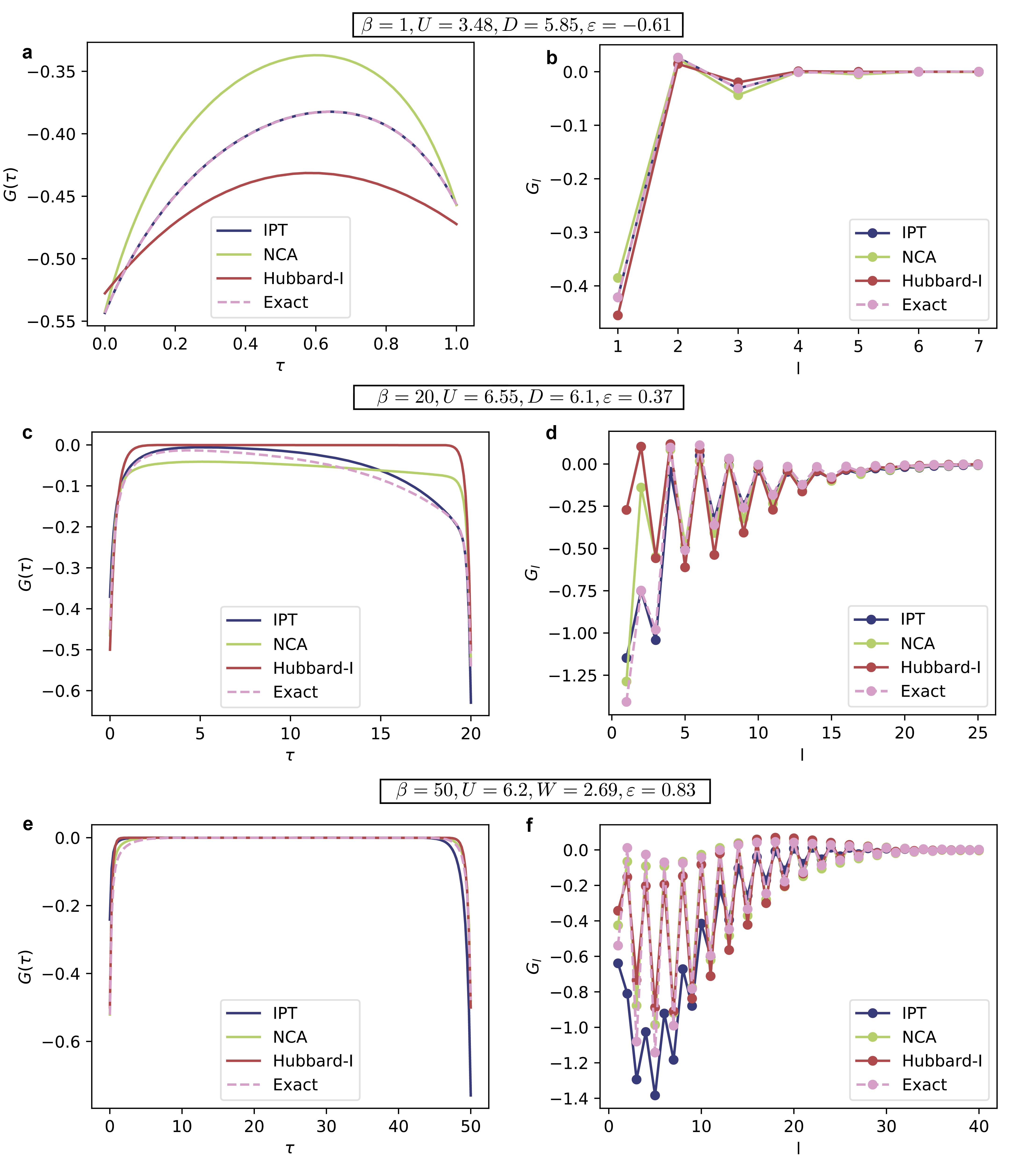}
  \caption[Randomly sampled configurations from Anderson Impurity Model database]{Representative randomly sampled configurations (from the database of
Table \ref{tab:database_1}) of the Anderson Impurity Model with a collection
of approximate exact solutions for the $\beta = 1 $ eV$^{-1}$ \textbf{a} imaginary-time
and \textbf{b} Legendre bases. Similarly for $\beta=20$ eV$^{-1}$ in the \textbf{c}
imaginary-time and \textbf{d} Legendre bases. Similarly for $\beta=50$ eV$^{-1}$ in the \textbf{e}
imaginary-time and \textbf{f} Legendre bases.}
  \label{fig:database_gfs}
\end{figure*}

In this section we provide representative test examples that are 
discussed in this manuscript for illustration of the methodology.
We first turn our attention to the representative inputs $\mathbf{x}_i$ and
outputs $\mathbf{y}_i$ of the training model (as per Fig. \ref{fig:database_cartoon}) we 
show the approximate solutions IPT  , NCA and Hubbard-I with the 
corresponding exact solution in both imaginary-time and Legendre bases for 
different inverse temperatures $\beta$ (as per Fig. \ref{fig:database_gfs}). 
The specific AIM configurations were chosen at random for the database
of $10,000$ solutions.

For $\beta=1$ eV$^{-1}$ and the parameter set $\{ U = 3.48 eV, W =
5.85 eV, \varepsilon = -0.61 \}$ we see that the IPT   approximation
is highly accurate in both bases (Fig. \ref{fig:database_gfs}a-b), even away from
half-filling. Moreover, whilst both Hubbard-I and NCA
solvers are quantitatively in disagreement with the exact solution, the NCA solver
recovers some critical features of the exact solution such as the
$G(\tau\rightarrow 0)$ value as well as the overall shape. Evidently,
the Legendre basis is significantly more compact than the imaginary-time basis,
however the error does not manifest itself so obviously in it, where small
changes in the coefficient can give large changes in the overall physical
properties.

For $\beta=20$ eV$^{-1}$ the parameter set $\{ U = 6.55 eV, W = 6.1 eV,
\varepsilon = 0.31 \}$ is randomly chosen instead, and illustrated in
Fig. \ref{fig:database_gfs}c-d. The IPT approximation in
this case retains a qualitative agreement with the exact answer, however there
are obvious deviations that emerge. Furthermore, the NCA and Hubbard-I solvers
are in quantitative and qualitative disagreement, and this is illustrated
in the second and third Legendre coefficients. For example, the Hubbard-I solver
indicates an insulating solution when in reality the system is metallic.

For $\beta=50$ eV$^{-1}$  the parameter $\{ U = 6.2 eV, W = 2.69 eV,
\varepsilon = 0.83 \}$ again randomly chosen, and illustrated in Fig. \ref{fig:database_gfs}e-f.
Here, in contrast with the higher temperature results, both Hubbard-I and NCA solvers are quantitatively 
and qualitatively in better agreement with the exact solution than the IPT result. Moreover, the IPT result 
qualitatively estimates an incorrect value for the onsite electron occupation, as estimated by $G(\tau\rightarrow0)$.

\subsection*{Appendix D: Neural Network loss functions for additional models}

\begin{figure*}
\includegraphics[width=1.7\columnwidth]{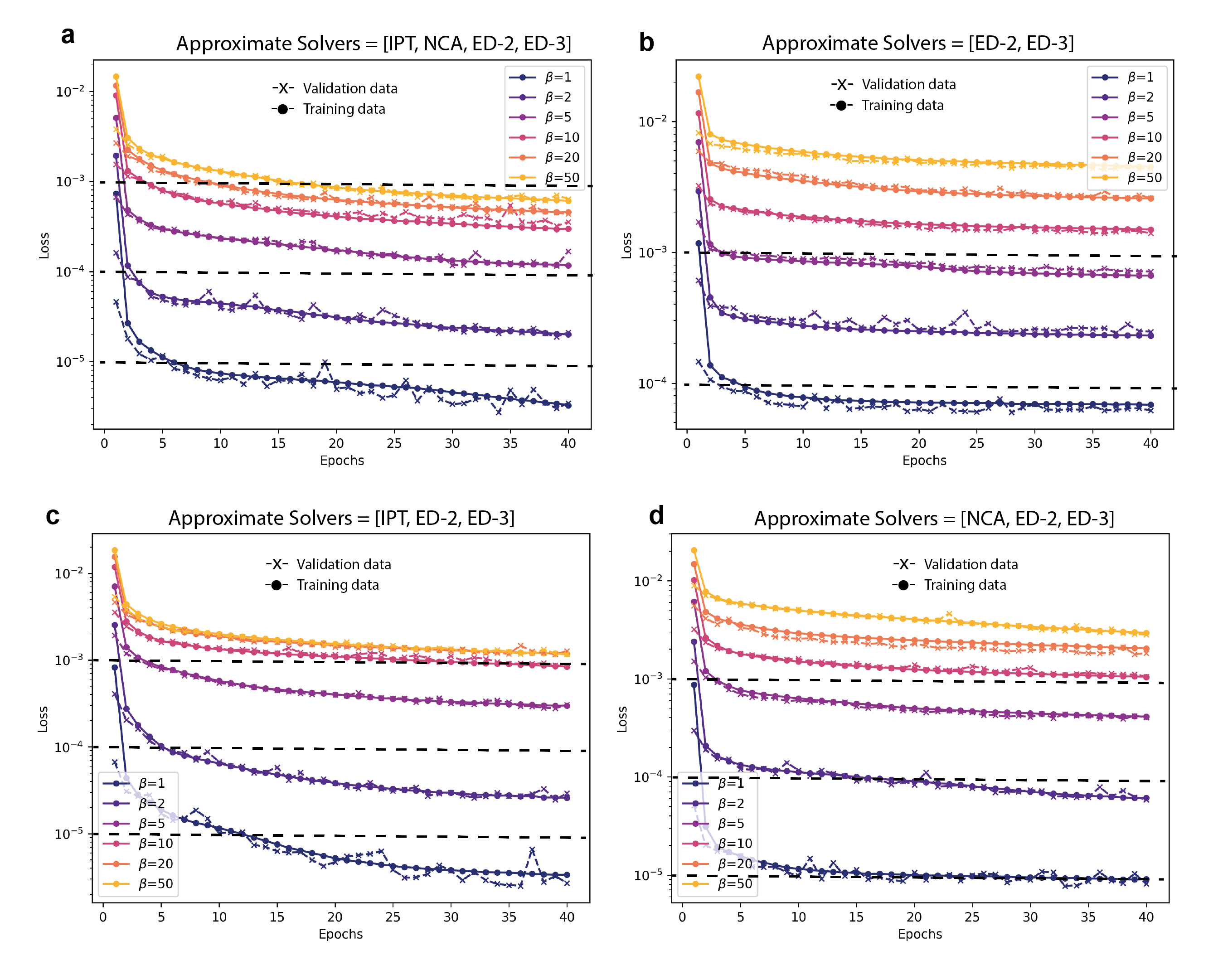}
  \caption{Cost function for the adapative $\tau$ mesh for different inverse temperatures $\beta$ using as input to the neural network the solvers \textbf{a} [IPT, NCA, ED-1, ED-2] \textbf{b} [ED-1, ED-2, ED-3, Hubbard-I], \textbf{c} [IPT, ED-1, ED-2], and \textbf{d} [NCA, ED-1, ED-2]. \edits{IPT is the Iterated Perturbation Theory Solver, NCA is the Non Crossing Approximation solver and ED-[1,2,3] are the truncated ED solvers.}}
  \label{fig:app_loss}
\end{figure*}

Figure~(\ref{fig:app_loss}) illustrates the validation loss for all other models studied in the main text. The overwhelming conclusion that can be made from these results is that the combination of quality approximate solutions results in smaller values of the validation loss when using the imaginary-time basis. As mentioned in the main text, using the IPT approximate solver is a critical  component to minimising the overall loss of the neural network. 

\subsection*{Appendix E: Neural Network prediction for representative samples}

\begin{figure*}[h!] \centering
\includegraphics[width=1.8\columnwidth]{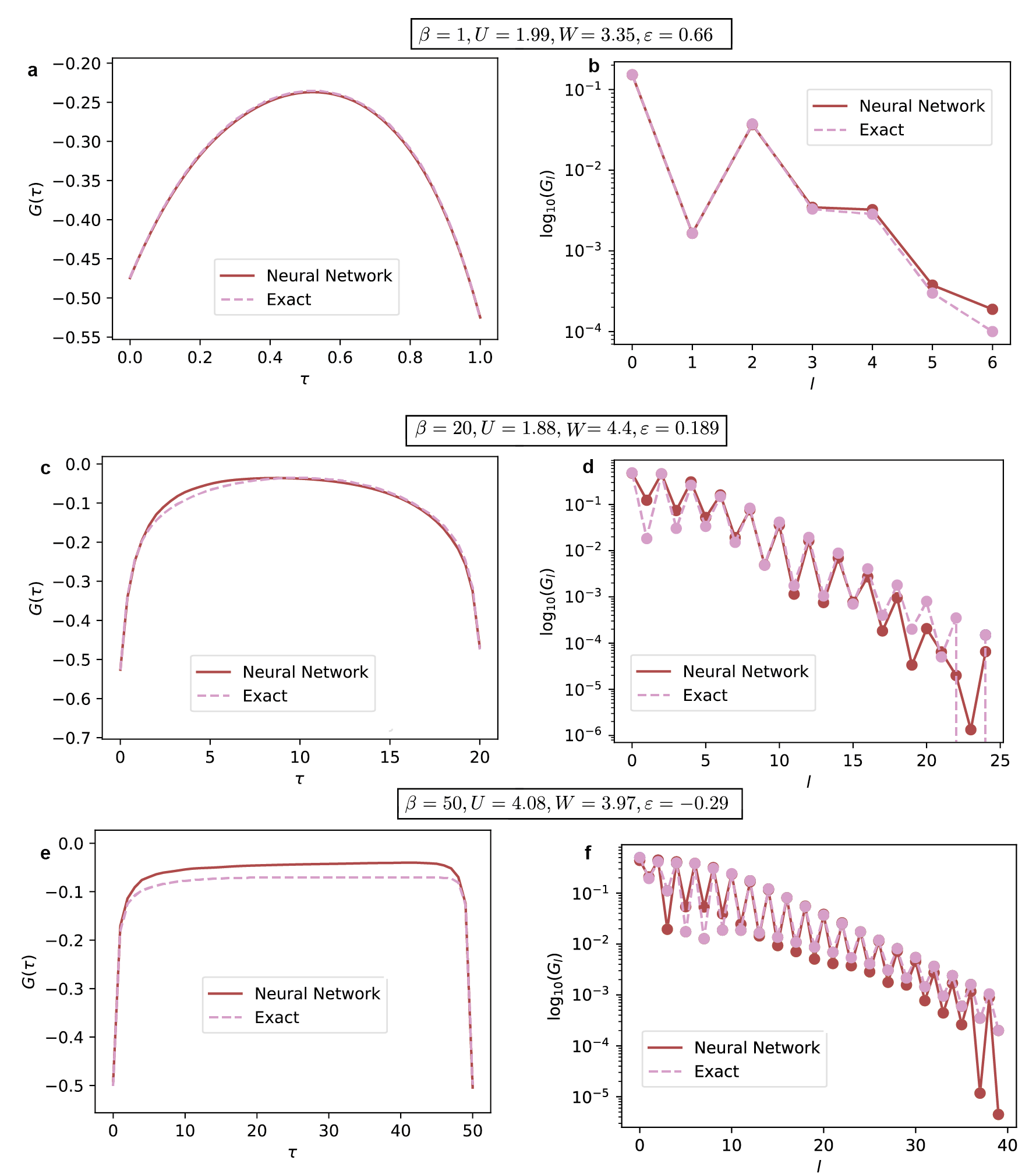}
  \caption[Predictions using the Neural Network for the Anderson Impurity Model]{Representative randomly sampled configurations of the Anderson Impurity Model neural network solution and exact
solution for the $\beta = 1 $ eV$^{-1}$ trained on the $\mathcal{M}_1$ ensemble of approximate solutions in the \textbf{a} imaginary-time and \textbf{b} Legendre
bases. Similarly for $\beta=20$ eV$^{-1}$ in the \textbf{c} imaginary-time and
\textbf{d} Legendre bases. Finally, for $\beta=50$ eV$^{-1}$ in the \textbf{f} imaginary-time and
\textbf{f} Legendre bases.}
  \label{fig:nn_predictions}
\end{figure*}

Having optimised the various aspects of the network, and analysed its loss
functions, we discuss here how the model is able to make predictions for unseen parameter 
choices (see Fig. \ref{fig:nn_predictions}). For $\beta = 1$ eV$^{-1}$ and the parameter set $\{ U =
1.99 eV, W = 3.35 eV, \varepsilon = 0.66 \}$ we see that the neural network captures
the essence of the error in both the imaginary-time and Legendre bases
(Fig. \ref{fig:nn_predictions}a-b). This is perhaps most clearly identified
at $G(\tau=0.6)$ in the imaginary-time basis or at second Legendre coefficient
in the polynomial basis. For the Legendre basis, we present the data on a
log-scale to highlight how the prediction is essentially error-free until 
the coefficients become $<10^{-3}$ in log-space, which is practically negligible
for the reconstruction in imaginary-time. On the other hand, for $\beta=20$ eV$^{-1}$
and the parameter set $\{ U = 1.88 eV, W = 4.4 eV, \varepsilon = 0.189 \}$ we see that
(Fig. \ref{fig:nn_predictions}c-d), while the agreement between the neural
network solver is not as refined compared to the $\beta=1$ eV$^{-1}$ network, it
is much better than what the approximate solvers predict. Interestingly, it is
only through the odd-components of $G_l$ that the network actually struggles
with predicting (up to a critical value of $l$), which suggests that separating
these contributions (\emph{i.e.} the half-filling and away from half-filling) and
learning them separately could improve the network. By extending into the lower 
temperature regime, \emph{i.e} $\beta=50$ eV$^{-1}$, we see that similar features 
persist in the even/odd decomposition of the coefficients, illustrated for the the parameter set $\{ U = 1.88 eV, W = 4.4 eV, \varepsilon = 0.189 \}$ in (Fig. \ref{fig:nn_predictions}e-f). Again, in contrast to the $\beta=1$ eV$^{-1}$ results, we see that there are slight discrepancies throughout the entire range of Legendre coefficients, rather 
than just the higher-order ones. Nevertheless, the agreement in the imaginary-time basis indicates that many of the main physically relevent features are captured by the neural network up to $\beta=50$eV$^{-1}$, such as the occupation and metallicity of the solution.

\end{document}